%% file: 2006.03055.tex
\documentclass[a4paper,11pt]{article}

\usepackage{jheppub}
\usepackage{graphicx}
\usepackage{amsmath}
\usepackage{slashed}
\usepackage{braket}
\usepackage[export]{adjustbox}
\usepackage{enumitem}
\usepackage{placeins}
\usepackage[normalem]{ulem}

\newcommand{\refcite}[1]{ref.~\cite{#1}}
\newcommand{\refscite}[1]{refs.~\cite{#1}}
\newcommand{\Eq}[1]{Eq.~\eqref{eq:#1}}
\newcommand{\eq}[1]{eq.~\eqref{eq:#1}}
\newcommand{\eqs}[2]{eqs.~\eqref{eq:#1} and \eqref{eq:#2}}

\renewcommand{\sec}[1]{section~\ref{sec:#1}}
\newcommand{\secs}[2]{sections~\ref{sec:#1} and \ref{sec:#2}}

\newcommand{\df}{\mathrm{d}}
\newcommand{\img}{\mathrm{i}}
\newcommand{\eps}{\epsilon}

\newcommand{\bn}{{\bar n}}
\newcommand{\bq}{{\bar q}}

\newcommand{\wa}{{w_1}}
\newcommand{\wb}{{w_2}}
\newcommand{\Obs}{\Tau}

\newcommand{\cG}{\mathcal{G}}
\newcommand{\cI}{\mathcal{I}}

\newcommand{\cL}{\mathcal{L}}
\newcommand{\cM}{\mathcal{M}}
\newcommand{\cN}{\mathcal{N}}
\newcommand{\cO}{\mathcal{O}}
\newcommand{\Tau}{\mathcal{T}}
\newcommand{\cP}{\mathcal{P}}

\newcommand{\qt}{{\vec q}_T}
\newcommand{\bt}{{\vec b}_T}

\newcommand{\nlim}{\lim\limits_{n-\text{coll.}}}
\newcommand{\slim}{\lim\limits_{s-\text{coll.}}}
\newcommand{\strictlim}{\lim\limits_{\text{strict }n-\text{coll.}}}
\newcommand{\as}{\alpha_s}

\newcommand{\nn}{\nonumber}
\newcommand{\MSbar}{\overline{\mathrm{MS}}}
\newcommand{\lqcd}{\Lambda_\mathrm{QCD}}

\def\beq{\begin{equation}}
\def\eeq{\end{equation}}
\def\bea{\begin{eqnarray}}
\def\eea{\end{eqnarray}}

\title{Collinear expansion for color singlet cross sections}

\author[a]{Markus A.~Ebert,}
\emailAdd{ebert@mit.edu}

\author[b]{Bernhard Mistlberger,}
\emailAdd{bernhard.mistlberger@gmail.com}

\author[a]{and Gherardo Vita}
\emailAdd{vita@mit.edu}

\affiliation[a]{Center for Theoretical Physics, Massachusetts Institute of Technology, Cambridge, Massachusetts 02139, USA}
\affiliation[b]{SLAC National Accelerator Laboratory, Stanford University, Stanford, CA 94039, USA}

\abstract{
We demonstrate how to efficiently expand cross sections for color-singlet production at hadron colliders
around the kinematic limit of all final state radiation being collinear to one of the incoming hadrons.
This expansion is systematically improvable and applicable to a large class of physical observables.
We demonstrate the viability of this technique by obtaining the first two terms in the collinear expansion of the rapidity distribution of the gluon fusion Higgs boson production cross section at next-to-next-to leading order (NNLO) in QCD perturbation theory.
Furthermore, we illustrate how this technique is used to extract universal building blocks of scattering cross section like the $N$-jettiness and transverse momentum beam function at NNLO.
}
\preprint{MIT-CTP 5207, SLAC-PUB-17536}

\begin{document}

\maketitle

\input{Chapters/Introduction.tex}

\input{Chapters/Setup.tex}

\input{Chapters/ExpansionCrossSection.tex}

\input{Chapters/ExpansionMatrixElements.tex}

\input{Chapters/SCET.tex}
\input{Chapters/BeamFunctions.tex}

\input{Chapters/Rapidity.tex}

\input{Chapters/Conclusion.tex}

\bibliographystyle{jhep}
\bibliography{../refs}

\end{document}

%% file: Chapters/Introduction.tex
\section{Introduction}
Our knowledge of the structure of quantum field theory (QFT) is rapidly advancing. 
On the one hand this steady progress allows us to answer fundamental questions about the interactions of nature by  deriving precise predictions for the outcome of scattering experiments that can be compared with experimental observation.
On the other hand we learn about the mathematical structures that underly this description.

Progress in QCD perturbation theory has allowed us to venture to predictions at next-to-next-to-next-to leading order (N$^3$LO) in the strong coupling constant for select inclusive and differential cross sections at the Large Hadron Collider (LHC)~\cite{Anastasiou:2015ema,Dreyer:2016oyx,Mistlberger:2018etf,Cieri:2018oms,Dulat:2018bfe,Dreyer:2018qbw,Duhr:2019kwi,Duhr:2020seh,Duhr:2020kzd}.
Resummation of kinematic limits of cross sections has reached the similarly astounding precision for a multitude of observables~\cite{Becher:2008cf, Abbate:2010xh, Hoang:2015hka, Bonvini:2014joa, Schmidt:2015cea, H:2019dcl, Ebert:2017uel, Chen:2018pzu, Bizon:2018foh}.
Nevertheless, the difficulty of describing the scattering of fundamental particles is ever rising with increasing demand for precision and for more complex observables. 
To overcome seemingly insurmountable complexity, parametric or kinematic expansions have proven highly effective.
For example, expanding the gluon fusion Higgs boson production cross section around the production threshold of the Higgs boson allowed for the computation of the first hadron collider cross section at N$^3$LO in QCD perturbation theory~\cite{Anastasiou:2015ema}.

Kinematic expansions in hadron collisions have been studied since a long time. For example, such expansions provide the bases of factorization theorems for inclusive processes in hadron collisions~\cite{Bodwin:1984hc,Collins:1984kg,Collins:1985ue,Sterman:1986aj,Collins:1988ig,Collins:1989gx,Catani:1989ne,Lustermans:2019cau}.
They have also been used to derive universal quantities like emission currents or splitting amplitudes (see for example \refscite{Altarelli:1977zs,DelDuca:1989jt,Mangano:1990by,Bern:1994zx,Campbell:1997hg,Catani:1998nv,Catani:2000vq, deFlorian:2001zd,Becher:2009cu,Dixon:2010zz,Catani:2010pd,Duhr:2013msa,Duhr:2014nda,Li:2014afw,Anastasiou:2018fjr,DelDuca:2019ggv,Dixon:2019lnw}), for studying the high energy behavior of amplitudes and cross sections (see for example \refscite{Kuraev:1977fs,Balitsky:1978ic,Lipatov:1985uk,Catani:1990eg,Mueller:1994jq,Korchemskaya:1996je,DelDuca:2001gu,Caron-Huot:2017zfo,Caron-Huot:2020grv}) as well as in the calculation of counterterms for subtraction algorithms (see for example~\refscite{Gehrmann_De_Ridder_2012,DelDuca:2013kw,Czakon:2014oma,Herzog:2018ily,Magnea:2018hab,Delto:2019asp}).
In the method of regions, one expands Feynman integrals in all relevant kinematic limits to simplify their evaluation~\cite{Beneke:1997zp}.
More generally they can be used to study divergence structures of Feynman integrals~\cite{Anastasiou:2018rib} or to approximate hadronic cross sections~\cite{Dittmaier:2014qza,Bonetti:2017ovy,Lindert:2018iug,Anastasiou:2018adr,Liu:2019tuy,Dreyer:2020urf}.
Soft-Collinear Effective Theory (SCET) is based on the kinematic expansion of scattering amplitudes and the realisation that such limits can be described by effective field theories~\cite{Bauer:2000ew, Bauer:2000yr, Bauer:2001ct, Bauer:2001yt, Bauer:2002nz}.
These techniques have also been used to derive the factorization of several infrared observables for color-singlet processes at hadron colliders, see for example \refscite{Stewart:2009yx,Becher:2010tm,GarciaEchevarria:2011rb,Chiu:2012ir,Monni:2016ktx,Buffing:2017mqm,Tackmann:2012bt,Banfi:2012jm,Becher:2012qa,Liu:2012sz,Procura:2014cba,Lustermans:2019plv,Monni:2019yyr}.

In this article we detail a technique for the efficient expansion of differential partonic cross sections for the production of a color singlet final state $h$ in hadron-hadron collisions in the kinematic limit that all radiation produced alongside $h$ is collinear to one of the collision axis of our scattering process.
The method outlined here is based on the work mentioned before and extends existing technology.
It also shares many similarities with the method developed in \refscite{Anastasiou:2015yha,Anastasiou:2013srw,Anastasiou:2014lda,Anastasiou:2014vaa,Dulat:2017prg} to expand cross sections around the limit of all radiation being soft.
Our expansion is carried out at the integrand level, i.e., before loop or phase space integrals are carried out. 
The resulting expressions can be interpreted diagrammatically.
This in turn greatly simplifies the analytic computation of matrix elements by employing powerful loop integration techniques like the reverse unitarity framework~\cite{Anastasiou2003,Anastasiou:2002qz,Anastasiou:2003yy,Anastasiou2005,Anastasiou2004a} or integration-by-part (IBP) identities~\cite{Chetyrkin:1981qh,Tkachov:1981wb}.
Our expansion is systematically improvable as we can compute to arbitrarily high power in our expansion parameter. 
The mathematical functions that appear in each term of the expansion are determined by the first few expansion coefficients. 

The collinear expansion of cross sections can find many applications in the computation of higher order corrections to scattering processes.
Cross sections for the production of hard probes $h$ can be approximated by performing a systematic collinear expansion.
Recently, an all-order factorization theorem was derived for the first order in this collinear expansion~\cite{Lustermans:2019cau}.
While the usefulness of our expansion technique depends on the specific observable in question,
it is obvious that key observables like the rapidity or transverse momentum of a hard probe are amenable to such an expansion.
We demonstrate the applicability of our collinear expansion to the rapidity distribution of the Higgs boson produced via gluon fusion. By calculating the collinear expansion to its second order, we demonstrate the excellent convergence of our series towards the full result at NNLO in perturbation theory.

Kinematic limits of cross sections can also be used to identify universal structures of quantum field theories.
Our expansion technique allows to gain access to splitting functions or integrated counter terms that may find application in subtraction algorithms used for the computation of fully differential cross sections. 
Universal building blocks that find their application in the resummation of perturbative cross sections can be accessed efficiently using this expansion technique. 

One example of such universal building blocks are so-called beam functions~\cite{Stewart:2009yx,Stewart:2010qs}
which arise in SCET and play a crucial role in factorization theorems of hadronic observables.
We demonstrate how to relate beam functions to the kinematic limit of our perturbative cross sections and how they can be extracted efficiently.
Specifically, we investigate the transverse momentum $(q_T)$ dependent beam functions and $N$-jettiness ($\Tau_N$) beam function.
We illustrate our method by computing these quantities through NNLO, up to the second order in the dimensional regularization parameter $\eps$, confirming recent results in the literature~\cite{Luo:2019hmp,Luo:2019bmw,Baranowski:2020xlp}.
These results are necessary input for the calculation of aforementioned beam functions at N$^3$LO in QCD,
where much progress has been already made for the quark $\Tau_N$ beam function~\cite{Melnikov:2018jxb,Melnikov:2019pdm,Behring:2019quf}, and which has already been achieved for the quark $q_T$ beam function and TMDPDF~\cite{Luo:2019szz}.
In our companion papers~\cite{Ebert:2020yqt,Ebert:2020unb}, we complete this task by computing the $q_T$ and $\Tau_0$ beam functions in all channels at N$^3$LO based on the methods outlined in this article.

In recent years the universal structure of cross sections beyond leading power in kinematic expansions within SCET have been explored \cite{Manohar:2002fd,Beneke:2002ph,Pirjol:2002km,Beneke:2002ni,Bauer:2003mga,Hill:2004if,Lee:2004ja,Benzke:2010js,Freedman:2014uta,Kolodrubetz:2016uim,Moult:2016fqy,Moult:2017jsg,Beneke:2017vpq,Feige:2017zci,Moult:2017rpl,Chang:2017atu,Moult:2017xpp,Alte:2018nbn,Beneke:2018gvs,Beneke:2017ztn,Beneke:2018rbh,Moult:2018jjd,Ebert:2018lzn,Ebert:2018gsn,Bhattacharya:2018vph,Beneke:2019kgv,Moult:2019mog,Beneke:2019mua,Moult:2019uhz,Moult:2019vou,Liu:2019oav,Liu:2020ydl,Liu:2020eqe}.
As this avenue of research is still growing rapidly, our expansion techniques may provide analytic information towards the structure of cross sections at higher power.  
In fact, the method developed in this paper is inspired by the calculation of power corrections in fixed order SCET for $\Tau_0$ \cite{Moult:2016fqy,Moult:2017jsg,Ebert:2018lzn} and $q_T$~\cite{Ebert:2018gsn}.
It will be interesting to extend these studies to higher order in $\as$ and the power expansion.
We hope that our techniques will provide readily accessible tools for the computation of yet unknown universal building blocks.

This article is structured as follows: In \sec{setup} we setup a parameterization for differential cross sections for color singlet production at hadron colliders.
This will mainly serve to develop a notation and to identify the objects that we aim to expand. 
In \sec{collinear_expansion_xs} we introduce the general strategy of expanding differential hadronic cross sections around the collinear limit, identifying the relevant kinematic regions and formally defining what we intend by collinear expansion.
We then continue the discussion about collinear expansions in \sec{collinear_expansion} by showing in practice how to perform the collinear expansion for squared matrix elements.
We will show explicit examples of the expansion of two loop cut diagrams at leading and beyond leading power, both for real radiation as well as for loop corrections.
In \sec{SCET} we explain how our collinear expansion of cross section is related to the effective field theory framework of SCET and in particular to the factorization of hadronic differential cross sections.
In \sec{beamfunctions} we review the role of SCET beam functions in the factorization of hadronic differential cross sections and we show that they are naturally connected to the leading term of our collinear expansion of cross sections. We discuss in detail how to obtain beam functions both in the case of $q_T$ and $\Tau_N$.
In \sec{rapidity} we apply our formalism to compute the rapidity spectrum of the Higgs in gluon fusion at NNLO in QCD via the collinear expansion of the partonic cross section.
We conclude in \sec{conclusions}.
Our two-loop results for the $q_T$ and $\Tau_N$ bare beam functions are provided as ancillary files.

%% file: Chapters/Setup.tex
%%%%%%%%%%%%%%%%%%%%%%%%%%%%%%%%%%%%%%%%%%%%%%%%%%%%%%%%%%%%%%%%%%%%%%%%%%%%%%%%
\section{Setup for differential cross sections}
\label{sec:setup}
%%%%%%%%%%%%%%%%%%%%%%%%%%%%%%%%%%%%%%%%%%%%%%%%%%%%%%%%%%%%%%%%%%%%%%%%%%%%%%%%

In this section, we develop the notation for differential cross sections at hadron colliders.
In \sec{setup_general}, we introduce our generic notation for the production of a colorless hard probe $h$ in a proton-proton collision.
In \sec{setup_phasespace} we provide a detailed derivation of the required differential phase space.

%===============================================================================
\subsection{General setup and notation}
\label{sec:setup_general}
%===============================================================================

We consider the production of a colorless hard probe $h$ and an additional hadronic state $X$ in a proton-proton collision.
Examples of such processes are the gluon fusion production cross section of a Higgs boson or the hadronic production of a $Z$ boson or virtual photon (Drell-Yan).
\begin{align} \label{eq:process_hadr}
 P(P_1) + P(P_2) \quad\to\quad h(-p_h) + X(-k)
\,.\end{align}
Here, $P_{1,2}$ are the momenta of the incoming protons, which in the hadronic center-of-mass frame are given by
\begin{align} \label{eq:P1P2}
 P_1^\mu = \sqrt{S} \frac{n^\mu}{2} \,,\qquad P_2^\mu = \sqrt{S} \frac{\bn^\mu}{2}
\,,\end{align}
where $S = (P_1 + P_2)^2$ is the hadronic center-of-mass energy and the protons are aligned along the directions
\begin{align} \label{eq:nnb}
 n^\mu = (1,0,0,1) \,,\qquad \bn^\mu = (1,0,0,-1)
\,.\end{align}
In \eq{process_hadr}, $p_h$ is the momentum of the hard probe $h$, and $k$ is the total momentum of the hadronic state $X$,
and as indicated both momenta are taken to be incoming.

The hadronic process in \eq{process_hadr} receives contributions from the partonic processes
\begin{align} \label{eq:process_part}
 i(p_1) + j(p_2) \quad\to\quad h(-p_h) + X_n(-p_3, \dots, -p_{n+2})
\,,\end{align}
where $i$ and $j$ are the flavors of the incoming partons, and their momenta are given by
\begin{align} \label{eq:p1p2}
 p_1^\mu = x_1 P_1^\mu \,,\qquad p_2^\mu = x_2 P_2^\mu
\,,\end{align}
such that the partonic center of mass energy is given by
\begin{align}
 s = (p_1 + p_2)^2 = x_1 x_2 S
\,.\end{align}
In \eq{process_part}, $X_n$ is a hadronic final state consisting of $n\ge0$ partons with momenta $\{p_3, \cdots, p_{n+2}\}$ and total momentum $k^\mu \equiv \sum_{i>2} p_i^\mu$. 

We are interested in describing processes that are differential in the four momentum $p_h^\mu$,
which we parameterize in terms of its rapidity $Y$ and virtuality $Q$,
\beq
\label{eq:hadrvardef}
Y=\frac{1}{2}\log\left(\frac{\bar n\cdot p_h}{n\cdot p_h}\right),\hspace{1cm} Q^2=p_h^2\,,
\eeq
and by momentum conservation its transverse momentum $p_{h \perp}^\mu$ is fixed to be $p_{h \perp}^\mu = -k_\perp^\mu$.
The momentum $k^\mu$ is parameterized in terms of the variables
\bea
\label{eq:vardef}
\wa=-\frac{\bar n\cdot k}{\bar n \cdot p_1},\hspace{1cm}
\wb=-\frac{ n\cdot k}{ n \cdot p_2},\hspace{1cm}
x=\frac{k^2}{(\bar n\cdot k)(n\cdot k)} = 1- \frac{\vec{k}_\perp^2}{(\bar n\cdot k)(n\cdot k)} \,.
\eea

We refer to the hadronic cross section differential in the above variables as the general differential cross section,
\beq
\label{eq:sigma_differential}
 \frac{\df\sigma}{\df Q^2 \df Y \df \wa \df \wb \df  x}  =  \frac{\sigma_0}{\tau} \sum_{i,j} x_1 f_i\left(x_1\right) x_2 f_j\left(x_2\right) \frac{\df \eta_{ij}}{ \df Q^2  \df \wa \df \wb \df  x}  \,.
\eeq
Here, the sum runs over all possible initial state configurations $i,j$, the $f_i(x)$ denote the parton distribution functions, and $\df \eta_{ij} / (\df Q^2  \df \wa \df \wb \df  x)$ is the general partonic coefficient function.
\Eq{sigma_differential} is normalized by $\sigma_0$, which contains all constant factors appearing in the Born level cross section.
The Bjorken momentum fractions $x_{1,2}$ can be expressed in terms of the variables introduced above.
\bea
\label{eq:xidef}
x_1&=& \frac{x_1^B} {z_1}    =x_1^B  \biggl[\sqrt{1+(k_T/Q)^2} - \frac{\bn \cdot k}{Q} e^{-Y} \biggr]
\,,\nn\\
x_2&=&  \frac{x_2^B} {z_2}  = x_2^B \biggl[\sqrt{1+(k_T/Q)^2} - \frac{n \cdot k}{Q} e^{+Y} \biggr]
\,,\eea
where the momentum fractions appearing at Born level are given by
\beq \label{eq:xiBdef}
x_1^B =\sqrt{\tau} e^Y,\hspace{1cm}x_2^B =\sqrt{\tau} e^{-Y}
\,,\eeq
where $\tau=Q^2 / S$
and we use the functions 
\beq
\label{eq:zdef}
z_1=\sqrt{\frac{1-\wa}{1-\wb}} \sqrt{1-\wa-\wb+\wa\wb x},\hspace{0.5cm}z_2=\sqrt{\frac{1-\wb}{1-\wa}} \sqrt{1-\wa-\wb+\wa\wb x}.
\eeq
At Born level, $k^\mu=0$, such that the momentum fractions $x_{1,2}$ reduce to $x_{1,2}^B$,
while in the presence of real radiation the kinematic constraint $k^\mu < 0$ dictates that $x_{1,2} \ge x_{1,2}^B$.

The general partonic coefficient function in \eq{sigma_differential} is given by
\begin{align} 
\label{eq:sigma_part}
  \frac{\df \eta_{ij}}{ \df Q^2  \df \wa \df \wb \df  x}  &
 = \frac{\tau}{\sigma_0} \frac{\cN_{ij}}{2 Q^2} \sum_{X_n}\int  \frac{\df\Phi_{h+n} }{ \df \wa \df \wb \df  x}\, |\cM_{ij\to h+X_n}|^2
\,.\end{align}
Here, the sum runs over all hadronic final states $X_n$ consisting of $n$ partons,
and $\df\Phi_{h+n}$ is the phase space measure of the $h+X_n$ final state which will be discussed in more detail in \sec{setup_phasespace}.
$|\cM_{ij\to h+X_n}|^2$ is the associated squared matrix element summed over final and initial state colors and helicities.
We have also pulled out the overall normalization factor $\cN_{ij}$, related to the spins and polarizations of the incoming partons. 
Depending on the initial state, it is given by
\begin{align}
 \cN_{gg} &= \frac{1}{4(n_c^2-1)^2(1-\epsilon)^2}
\,,\nn\\
 \cN_{qg} = \cN_{gq} &= \frac{1}{4(n_c^2-1) n_c(1-\epsilon)}
\,,\nn\\
 \cN_{qq} = \cN_{q\bq} = \cN_{qq'} = \cN_{q\bq'} &= \frac{1}{4n_c^2}
\,.\end{align}
Here, $g$, $q$ ($\bq)$ and $q'$ ($\bq'$) indicate a gluon, (anti-)quark, and (anti-)quark of different flavor than $q$, respectively.

We expand the general partonic coefficient function in $\as$ as
\begin{align} \label{eq:eta_ij_1}
 &\frac{\df\eta_{ij} }{\df Q^2 \df w_1 \df w_2 \df x}
 = \sum_{\ell=0}^\infty \left(\frac{\as}{\pi}\right)^{\ell}
   \frac{\df\eta_{ij}^{(\ell)}}{\df Q^2 \df w_1 \df w_2 \df x}
\\\nn&
 = \eta_{ij}^V \delta(w_1)\delta(w_2)\delta(x)
 \,+\, \sum_{\ell=1}^\infty \left(\frac{\as}{\pi}\right)^{\ell}
   \sum_{n,m=1}^\ell w_1^{-1-m\eps} w_2^{-1-n\eps}
   \frac{\df\eta_{ij}^{(\ell,m,n)}(w_1,w_2,x,Q^2)}{\df Q^2 \df w_1 \df w_2 \df x}
\,.\end{align}
Here, $\eta_{ij}^V$ contains the Born cross section and purely virtual corrections,
and can itself be expanded in $\as/\pi$ with the first term $\eta_{ij}^{V\,(0)}=\delta_{\bar{i}j}$ for flavour diagonal processes like Drell-Yan or Higgs production.
The $\eta_{ij}^{(\ell,m,n)}$ are separately holomorphic in the vicinity of $w_1 = 0$ or $w_2 = 0$.

The differential cross section for a specific observable $\Obs$ that only depends on $p_h^\mu$ and $k^\mu$ is obtained from our general differential cross section given in \eq{sigma_differential} as
\beq
\label{eq:sigma_hadr}
\frac{\df \sigma}{\df Q^2 \df Y \df\Obs} =
 \sigma_0 \sum_{i,j} f_i(x_1^B) \otimes_{x_1^B} \frac{\df \eta_{ij}(x_1^B,x_2^B)}{\df Q^2 \df Y \df\Obs}   \otimes_{x_2^B} f_j(x_2^B).
\eeq
Here, the convolution integral is defined as
\beq
\label{eq:convdef}
f(x) \otimes_x g(x)  = \int_x^1 \frac{\df z}{z} \,f(z) g\left(\frac{x}{z}\right)
\,.\eeq
The corresponding partonic coefficient function differential in $\Obs$ is given by
\bea
\label{eq:partcoef_special}
\frac{\df \eta_{ij}(y_1,y_2)}{\df Q^2 \df Y \df\Obs} &=& \int_0^1 \df x \int_0^\infty \df \wa\df \wb \,
\delta\left(y_1-z_1\right)  \delta\left(y_2-z_2\right)
\nn\\&&\times\,
\delta\bigl[\Obs-\Obs(Q,Y,\wa,\wb,x)\bigr] \, \frac{\df\eta_{ij}}{ \df Q^2  \df \wa \df \wb \df  x}
\,,\eea
where $\Obs(Q,Y,\wa,\wb,x)$ picks out the value of the observable at a given phase space point.
Note that in the above equation the variables $z_i$ are still functions of $\wa$, $\wb$ and $x$ as specified in \eq{zdef}.

The partonic coefficient function in~\eqref{eq:sigma_hadr} contains ultraviolet (UV) and infrared (IR) divergences.
We regulate such divergences using conventional dimensional regularisation by extending the space time dimension by an infinitesimal amount to be $d=4-2\epsilon$.
UV divergences are removed by renormalization in the $\MSbar$ scheme.
IR singularities are removed by the standard mass factorization redefinition of the PDFs.
Specifically, the unsubtracted PDF $f_i(x)$ is given in terms of the finite PDF in the $\MSbar$ scheme $f^R_i(x)$ as
\begin{align} \label{eq:pdf_renorm}
 f_i(x) &= \sum_j \Gamma_{ij}(z) \otimes_z f_j^R(z)
\,,\end{align}
where the sum runs over all parton flavors $j$, $\Gamma_{ij}$ is the PDF counterterm that is known through three loops~\cite{Moch:2004pa, Vogt:2004mw}, and we suppress the associated factorization scale $\mu$.
This allows us to write the hadronic differential cross section of \eq{sigma_hadr} in terms of finite quantities,
\beq
\label{eq:sigma_hadr_finite}
\frac{\df \sigma}{\df Q^2 \df Y \df\Obs} =
\sigma_0 \sum_{i,j} f_i^R(x_1^B) \otimes_{x_1^B} \frac{\df \eta_{ij}^R(x_1^B,x_2^B)}{\df Q^2 \df Y \df\Obs}   \otimes_{x_2^B} f_j^R(x_2^B)\,,
\eeq
with
\beq
 \frac{\df \eta_{ij}^R(z_1,z_2)}{\df Q^2 \df Y \df\Obs}=\sum_{k,\ell} \Gamma_{ki}(z_1) \otimes_{z_1}  \frac{\df \eta_{k\ell}(z_1,z_2)}{\df Q^2 \df Y \df\Obs}\otimes_{z_2} \Gamma_{\ell j}(z_2)\,.
\eeq

 %===============================================================================
\subsection{Differential phase space}
\label{sec:setup_phasespace}
%===============================================================================

To derive the phase space differential in the variables defined in \eq{vardef}, we start from the generic expression for the phasespace of the $h+X_n$ system,
\begin{align} 
\label{eq:measure}
 \df\Phi_{h+n} &=
  \frac{\df^dp_h}{(2\pi)^d} (2\pi)\delta_+(p_h^2-Q^2) \,
  \left[\prod\limits_{i=3}^{n+2}\frac{\df^dp_i}{(2\pi)^d} (2\pi)\delta_+(p_i^2)\right]
   (2\pi)^d \delta^d(p_1+p_2+p_h+k)
\,,\end{align}
where
\beq
 \delta_+(p^2-m^2)=\theta(-p^0-m)\delta(p^2-m^2)
\,,\eeq
and $k^\mu= \sum\limits_{i=3}^{n+2} p_i^\mu$ is the total momentum of $X_n$.
Next, we separate the integration over $p_h$ and $k$ by inserting the unity
\beq
1=\int \frac{\df^dk}{(2\pi)^d} (2\pi)^d\delta^d(k-p_3-\dots-p_{n+2})\int_0^\infty \frac{\df\mu^2}{2\pi}(2\pi)\delta_+(k^2-\mu^2)\,.
\eeq
This splits the $h{+}X_n$ phase space measure into an integral over the phase space $\Phi_2^\text{m}$ for two massive particles
and the phase space $\Phi_n^0$ for $n$ massless partons of total invariant mass $\mu^2$,
\begin{align} \label{eq:Phi_hn}
 \df\Phi_{h+n} = \int_0^{\infty} \frac{\df\mu^2}{2\pi} \df\Phi_2^\text{m}(\mu^2) \, \df\Phi_n^0(\mu^2)
\,.\end{align}
The two phase space measures are defined as
\begin{align} \label{eq:phi_2_and_0}
 \df\Phi_2^\mathrm{m}(\mu^2) &=
 \frac{\df^d p_h}{(2\pi)^d}\,(2\pi)\delta_+(p_h^2-Q^2) \frac{\df^d k}{(2\pi)^d} (2\pi)\delta_+(k^2-\mu^2)\,(2\pi)^d\delta^d(p_1+p_2+p_h+k)
\,,\nn\\
 \df\Phi^0_n(\mu^2) &=
 \biggl[\prod\limits_{i=3}^{n+2}\frac{\df^dp_i}{(2\pi)^d} (2\pi)\delta_+(p_i^2)\biggr]
 (2\pi)^d \delta^d\biggl(k-\sum\limits_{i=3}^{n+2} p_i\biggr)
\,.\end{align}
The on-shell constraint for $p_h$ is used together with the definition of the rapidity of \eq{hadrvardef} to define the born momentum fractions $x_{1,2}^B$.
Transforming from $k^\mu$ to the variables introduced in \eq{vardef}, we obtain the desired result for the differential phase space,
\begin{align} \label{eq:dPhi_wa_wb_x}
 \frac{\df\Phi_{h+n}}{\df \wa\df \wb \df x} &
 = \frac{\left(\wa\wb s\right)^{1-\epsilon} (1-x)^{-\epsilon}}{(4\pi)^{2-\epsilon} \Gamma(1-\epsilon)}
 \, \theta[x(1-x)] \, \theta(\wa) \, \theta(\wb) \, \df\Phi_n^0\left(s \wa\wb x\right)
\,.\end{align}
In the special case of having zero or one final state parton, \eq{dPhi_wa_wb_x} becomes
\begin{align}
 \frac{\df\Phi_{h+0}}{\df \wa\df \wb \df x} &= \frac{(2\pi)}{s} \delta(x) \delta(\wa)\delta(\wb)
\,,\nn\\
 \frac{\df\Phi_{h+1}}{\df \wa\df \wb \df x} &= \frac{(\wa\wb s)^{-\eps}}{2 (4\pi)^{1-\epsilon} \Gamma(1-\epsilon)} \delta(x) \theta(\wa)\theta(\wb)
\,.\end{align}
The inclusive phase space volume is obtained by integrating over the differential phase space volume,
\beq
\Phi_{h+n}=\int\df\wa \df\wb \df x \,  \delta\left(1-w_1-w_2+w_1 w_2 x -Q^2/s\right)  \frac{\df\Phi_{h+n}}{\df \wa\df \wb \df x}  .
\eeq

%% file: Chapters/ExpansionCrossSection.tex
%%%%%%%%%%%%%%%%%%%%%%%%%%%%%%%%%%%%%%%%%%%%%%%%%%%%%%%%%%%%%%%%%%%%%%%%%%%%%%%%
\section{Collinear expansion of color-singlet cross sections}
\label{sec:collinear_expansion_xs}
%%%%%%%%%%%%%%%%%%%%%%%%%%%%%%%%%%%%%%%%%%%%%%%%%%%%%%%%%%%%%%%%%%%%%%%%%%%%%%%%

In this section, we introduce the general strategy of expanding cross sections around the collinear limit.
We begin by identifying the key kinematic regions in which we want to expand cross sections in \sec{expansions_intro}.
Next, we define the collinear expansion of hadronic cross sections in \sec{expansions_xsection}. 
Finally, we comment on the use of different coordinates in performing a collinear expansion in \sec{expandingvariables}.
We will provide explicit examples on how to implement this in practice for matrix elements in \sec{collinear_expansion}.

In this section, it will be very convenient to work with light-cone coordinates%
\footnote{Note that another popular conventions in the literature defines light-cone coordinates through the decomposition $p^\mu = p^- \frac{\bn^\mu}{\sqrt2} + p^+ \frac{n^\mu}{\sqrt2} + p_\perp^\mu$ with $p^\pm = (p^0 \pm p^z)/\sqrt2$.}.
We decompose a momentum $p^\mu$ as
\begin{align}\label{eq:lcdef}
 p^\mu &= p^+ \frac{\bn^\mu}{2} + p^- \frac{n^\mu}{2} + p_\perp^\mu \equiv (p^+, p^-, p_\perp)
\,,\end{align}
where the $p^\pm$ components are explicitly given by
\begin{align}\label{eq:lcdef2}
 p^- &= \bn \cdot p = p^0 + p^z \,,\quad p^+ = n \cdot p = p^0 - p^z
\,,\end{align}
and $p_\perp$ is the remaining transverse component.

%===============================================================================
\subsection{Power counting and modes}
\label{sec:expansions_intro}
%===============================================================================

Hadronic color singlet cross sections for infrared and collinear safe observables are finite quantities.
The perturbative description of such observables becomes inadequate when the value of the observable forces hadronic radiation produced on top of the colorless final state to be in the infrared or collinear regime.
For example, it is well known that when imposing an infrared-sensitive measurement $\Obs$, the cross section receives contributions of up to two logarithms $L = \ln(Q/\Obs)$ per order of the coupling constant, \emph{i.e.}\ $\sigma \sim \as^n L^{2n}$.
In the limit of $\Obs$ vanishing the presence of such logarithms hence signals the sensitivity to infrared physics and truncated perturbation theory does not yield an accurate description of the observable, as the large logarithms $L$ can overcome the suppression in $\as$.

In order to understand cross sections in their infrared and collinear kinematic regimes, it is necessary to identify and characterize these regimes.
Simply speaking, one can classify two particles with momenta $p_i$ and $p_j$ as collinear to each other when $p_i \cdot p_j \to 0$, while a particle with momentum $p_i$ is considered soft when $p_i^\mu \to 0$.
More precisely, particles should be classified as soft and collinear \emph{relative to the hard scale} of the process.
\begin{align} \label{eq:IR}
 p_i~\mathrm{collinear~to}~p_j \,:\quad p_i \cdot p_j \ll Q^2
\,,\qquad
 p_i~\mathrm{soft} \,:\quad p_i^\mu \ll Q
\,.\end{align}
In the case of sufficiently inclusive color-singlet processes, there are only two designated directions, namely the lightlike directions $n^\mu$ and $\bn^\mu$ of the incoming protons.
Employing the lightcone notation introduced in \eq{lcdef}, we can thus classify a momentum $p_i^\mu = (p^+, p^-, p_\perp)$ in the different dominant kinematic regions as
\begin{alignat}{3} 
\label{eq:modes}
 &\text{hard}:          \qquad &&p_i^\mu \sim Q \, (1, 1, 1)
\,,\nn\\
 &n\text{-collinear}:   \qquad &&p_i^\mu \sim Q \, (\lambda^2, 1, \lambda)
\,,\nn\\
 &\bn\text{-collinear}: \qquad &&p_i^\mu \sim Q \, (1, \lambda^2, \lambda)
\,,\nn\\
 &\text{soft}:          \qquad &&p_i^\mu \sim Q \, (\lambda^m, \lambda^m, \lambda^m)   \,,\qquad m=1,2
\,.\end{alignat}
Here, $\lambda \ll 1$ is an auxiliary power counting parameter indicating the suppression of the different modes relative to the hard scale $Q$.
Let us discuss \eq{modes} in more detail:
\begin{itemize}
 \item The hard region describes momenta directly associated with the production of the hard probe $h$.
       Since $h$ has invariant mass $p_h^2 = Q^2$, parametrically hard momenta also have virtuality $p_i^2 \sim Q^2$.
       For example, virtual corrections to the partonic process, \emph{i.e.}\ the form factor, are sensitive to this scaling.
 \item The $n$-collinear region describes a momentum where $n \cdot p_i \ll \bn \cdot p_i$, and hence $p_i$ is aligned with the $n$-direction.
       The scaling of the transverse component follows by noting that for on-shell particles $p^+_ip^-_i \sim p_{i,\perp}^2 \sim \lambda^2 Q^2$.
 \item The soft region describes low-energetic, but isotropic radiation, as is manifest from the homogeneous scaling in \eq{modes}.
       The choice of $m$ in \eq{modes} depends on whether the observable $\Obs$ under consideration is sensitive to the lightcone momenta only ($m=2$) or also to transverse momenta ($m=1$).
       In the SCET literature, these two cases are referred to as ultrasoft and soft, respectively, but in this work we simply refer to both cases as soft.
\end{itemize}
Note that in more general cases, such as also measuring final-state jets or complicated observables, more modes may arise, and this has given rise to a plethora of ``scaling hierarchies'' in the literature, see for example \refscite{Seymour:1997kj,Dasgupta:2001sh,Cheung:2009sg, Bauer:2011uc,vonManteuffel:2013vja,Larkoski:2014gra,Larkoski:2015zka, Larkoski:2015kga, Procura:2014cba,Becher:2015hka,Chien:2015cka,Ellis:2010rwa,Banfi:2010pa,Kelley:2012kj,Pietrulewicz:2016nwo,Hornig:2017pud,Neill:2018yet,Lee:2019lge,Chien:2019osu}.
For sufficiently inclusive observables as considered in this paper it suffices to only consider \eq{modes}.

Above, we have only given heuristic arguments for the observation that the scalings in \eq{modes} are the only relevant ones.
In fact, it is precisely the modes stated in \eq{modes} that arise in proofs of QCD factorization.
These proofs are based on the insight that there is a one-to-one correspondence between the momentum regions giving rise to the large $Q$ behavior and mass divergences in massless perturbation theory~\cite{Sterman:1978bi,Libby:1978bx}.
These mass divergences arise at pinch-singular surfaces, \emph{i.e.}~momentum regions when loop momenta can not be deformed away from singularities in the appearing propagators, and thus correspond to classically allowed scatterings.
The position of these singular surfaces can be determined using the Landau criteria \cite{Coleman:1965xm},
and one can then derive power counting rules for these singular surfaces to approximate amplitudes in the vicinity of the surface, and these rules precisely lead to the momentum regions shown in \eq{modes}.
For a more comprehensive discussion of these proofs, we refer to \refscite{Collins:1989gx,Sterman:1995fz,Collins:1350496}.
Similarly, these singular surfaces also arise when analysing Feynman diagrams using the method of regions~\cite{Beneke:1997zp}.
They are also the basis for the formulation of SCET~\cite{Bauer:2000ew, Bauer:2000yr, Bauer:2001ct, Bauer:2001yt, Bauer:2002nz}, an effective field theory to describe QCD in the infrared limit that separates quark and gluon fields into modes corresponding to \eq{modes}. We will discuss this connection in more detail in \sec{SCET}.

%===============================================================================
\subsection{Expanding cross sections around the collinear limit}
\label{sec:expansions_xsection}
%===============================================================================
Having characterized the relevant kinematic regions for infrared-sensitive observables,
we now discuss the expansion of hadronic cross sections of \eq{sigma_hadr} around the particular limit where all final state radiation becomes collinear to one of the incoming proton momenta.

Let us define our collinear expansion: we want to expand around the limit where all real momenta are treated as $n$-collinear, and thus the total momentum $k^\mu$ of the hadronic final-state is $n$-collinear as well, \emph{i.e.}\ it scales as
\begin{align} \label{eq:k_ncollinear}
 k^\mu \sim k^- \frac{n^\mu}{2} + \lambda^2 k^+ \frac{\bn^\mu}{2} + \lambda k_\perp^\mu
\,.\end{align}
We then want to expand the hadronic differential cross section in \eq{sigma_hadr} to obtain a power series in $\lambda$,
\beq
\frac{\df \sigma}{\df Q^2 \df Y \df\Obs} = \lambda^{-2} \frac{\df \sigma^{(0)}}{\df Q^2 \df Y \df\Obs}  + \lambda^{-1} \frac{\df \sigma^{(1)}}{\df Q^2 \df Y \df\Obs} +\dots.
\eeq
Here, the leading-power cross section $\sigma^{(0)}$ scales as $\lambda^{-2}$,%
\footnote{This leading-power collinear limit precisely corresponds to the generalized threshold limit of \refcite{Lustermans:2019cau}.}
the next-to-leading power (NLP) cross section%
\footnote{Note that for a large class of observables, as for example $q_T$ and beam thrust, the odd powers in this series vanish. It is therefore common practice to indicate as NLP the first non vanishing contribution beyond leading power, which in those cases would be $\sigma^{(2)}$.} 
$\sigma^{(1)}$ as $\lambda^{-1}$, and so forth.
Depending on the observable $\Obs$, this series may start at higher orders in $\lambda$,
but for the infrared-sensitive observables discussed in this paper we always encounter a leading $\cO(\lambda^{-2})$ term.

It is desirable that Born quantities like $Q^2$ and $Y$ are unaffected by the expansion we want to carry out, as they set the hard scales of the process.
The importance of this for expansions at subleading power in $\lambda$ was already stressed in \refscite{Ebert:2018gsn,Ebert:2018lzn}.
As a consequence, the Bjorken momentum fractions given in \eq{xidef} need to be expanded.
Expressing them in terms of hard quantities and the momentum $k$ we find
\begin{align} \label{eq:x12_expanded}
 \frac{x_1}{x_1^B} &=  \sqrt{1 + k_T^2 / Q^2} - \frac{k^-}{Q} e^{-Y}
      = 1 - \frac{k^- e^{-Y}}{Q} + \cO(\lambda^2)
\,,\nn\\
 \frac{x_2}{x_2^B} &= \sqrt{1 + k_T^2 / Q^2} - \frac{k^+}{Q} e^{+Y}
      = 1 + \cO(\lambda^2)
\,.\end{align}
Since the momentum fractions enter as arguments of the PDFs, a pure hadronic expansion to higher orders in $\lambda$ will automatically involve derivatives of PDFs, as firstly noted for $\Tau_0$ in~\refcite{Moult:2016fqy}.
Furthermore, the variables $\wa$ and $\wb$ we introduced in \eq{vardef} must also be expanded,
\beq
\label{eq:omexp}
\wa=\frac{-k^-}{x_1 \sqrt{S}}=\frac{-k^-}{x_1^B \sqrt{S}-k^-}+\mathcal{O}(\lambda^2)\hspace{1cm}\wb=\frac{-k^+}{x_2 \sqrt{S}}=\frac{-k^+}{x_2^B \sqrt{S}}+\mathcal{O}(\lambda^4)
\,,\eeq
where $x_{1,2}^B$ are the momentum fractions at Born level, see \eq{xiBdef}.
As a consequence we find that the $n$-collinear limit of \eq{partcoef_special} becomes
\bea
\label{eq:partoniccoefexp}
&&\nlim\frac{\df \eta_{ij}(y_1,y_2)}{\df Q^2 \df Y \df \Obs} =\delta\left(1-y_2\right)  \int_0^1 \df x \int_0^\infty \df \wa\df \wb \,   \delta\left[y_1-(1-\wa)\right]  \\
&&\hspace{0.5cm}\times \nlim \left\{\delta\left[\Obs-\Obs(Q,Y,\wa,\wb,x)\right] \frac{\df\eta_{ij}}{ \df Q^2  \df \wa \df \wb \df  x}\right\}\nn
\,,\eea
where $w_{1,2}$ are evaluated according to \eq{omexp}.
The definition of our observable $\Obs$ itself may not be invariant under rescaling according to our power counting.
In order to achieve a pure expansion of the hadronic cross section we may either expand the observable constraint or solve the constraint using one of the remaining integration variables and expand subsequently.
We address how the general partonic coefficient function $ \frac{\df\eta_{ij}}{ \df Q^2  \df \wa \df \wb \df  x}$ can be expanded in \sec{collinear_expansion}. 

Constructing a collinear expansion can be done with different objectives in mind. 
One objective can be to obtain a pure series expansion of the hadronic cross section as discussed above.
Another objective can be to simplify the computation of the partonic coefficient function which does not require a pure expansion of the hadronic cross section.  
In the latter scenario one would only expand the partonic coefficient function $\eta_{ij}$ on the right-hand side of \eq{partcoef_special}, but not expand the $w_{1,2}$ and the momentum fractions $x_{1,2}$ as presented above.
This approach can also serve as a suitable proxy to a collinear expansion, where parts of the cross section are kept exact.

%===============================================================================
\subsection{Expansions using different coordinates}
\label{sec:expandingvariables}
%===============================================================================

So far, we defined our power counting such that the invariant mass $Q^2$ and rapidity $Y$ of the produced hard probe $h$ scale homogeneously as $\cO(\lambda^0)$ and the lightcone components of the total momentum $k^\mu$ of the hadronic final state have a homogeneous power counting in $\lambda$. This is reasonable, since one can only measure directly the final-state particles in the hadronic collision, which are then used to define the power counting. In particular, $Q^2$ and $Y$ are the only hard scales in the considered hadronic cross section $\df\sigma/(\df Q^2 \df Y \df \Obs)$.

This setup immediately implies that the momenta $p_1$ and $p_2$ of the incoming partons do not have a homogeneous power counting, as it is evident from their explicit expressions in terms of $Q^2$, $Y$ and $k^\mu$,
\begin{align} \label{eq:p1p2_2}
 p_1^-(Q^2,Y,k^+,k^-,x)  &= -k^- + e^{+Y}\sqrt{Q^2 + k^+ k^- (1-x)}
\,,\nn\\
 p_2^+(Q^2,Y,k^+,k^-,x) &= -k^+ + e^{-Y}\sqrt{Q^2 + k^+ k^- (1-x)}
\,,\end{align}
see \eq{xidef}. Thus, $p_1$ and $p_2$ give rise to an infinite tower of power corrections in $\lambda$, which in turn requires an expansion of $\wa$ and $\wb$ used to define the general partonic coefficient function, as shown in \eq{omexp}.

Since one has access to all incoming and outgoing momenta in the calculation of the partonic coefficient function, the collinear expansion can also be defined by assigning a homogeneous power counting to $p_1$, $p_2$ and $k$. Since this assignment is only meaningful for the partonic process, we refer to it as \emph{partonic collinear expansion}. In this approach, the rescaling appropriate for the collinear limit is given by
\begin{align} \label{eq:partonic_expansion}
 p_1^- \to p_1^- \,,\quad p_2^+ \to p_2^+ \,,\quad \wa = -\frac{k^-}{p_1^-} \to \wa \,,\quad \wb = -\frac{k^+}{p_2^+} \to \lambda^2 \wb \,,\quad x \to x
\,.\end{align}
The key advantage of this assignment is that $\wa$ and $\wb$, which are the natural variables to express the partonic coefficient function, now have homogeneous power counting and do not need to be re-expanded themselves.
Thus, the collinear expansion has been reduced to an expansion in $\wb$.
The drawback of the partonic collinear expansion is that the rapidity $Y$ of hard probe $h$ no longer uniformly scales as $\cO(\lambda^0)$, which is evident from the expression
\begin{align}
 Y(p_1^-,p_2^+,k^+,k^-,x) &
 = \frac{1}{2}\log\left(\frac{p_1^- + k^-}{p_2^+ + k^+}\right)
 = \frac{1}{2}\log\left(\frac{p_1^- + k^-}{p_2^+}\right) + \cO(\lambda^2)
\,,\end{align}
which now is a quantity derived from $p_1^-$ and $p_2^+$, rather than fixing these as in \eq{p1p2_2}.

Comparing the rescalings in \eqs{omexp}{partonic_expansion}, we see that both approaches agree at leading power, but differ at subleading power. Since there is a well-defined relation between the two approaches, one can easily obtain one expansion from the other, but care has to be taken to consistently apply the power expansion.

In practice, each choice of defining the expansion has its advantages and disadvantages. We can discuss these by classifying the expansions according to the choice of independent variables used to express the partonic coefficient function, which by Lorentz invariance only requires four independent invariables.
It is useful to summarize the above observations for the following possibilities:
\begin{itemize}
 \item $(Q^2,Y,k^+,k^-,x)$: This parameterization has the advantage that is entirely expressed in terms of information about the final state momenta, including the Born measurements $Q$ and $Y$ of the hard probe $h$.
 As properties of the final state, $Q$ and $Y$ need not be expanded, and the collinear expansion is a strict expansion in $k^\mu$.
 Since partonic matrix elements are typically more concise when expressed in terms of the incoming momenta and the final-state radiation, the main drawback of this expansion is that it leads to lengthier expressions for the expanded matrix element.
 Furthermore, measuring the rapidity $Y$ fixes a reference frame for all momenta, such that boost invariance is not manifest anymore.
 \item $(Q^2, \wa, \wb, x)$: Since $\wa$ and $\wb$ are defined as ratios of Lorentz scalars, boost invariance is manifest in this parameterization. Its disadvantage is that $\wa$ and $\wb$ do not have manifest power counting in terms of the observables $Q^2$ and $Y$, and instead must be expanded in $k^\mu$ according to their definitions in \eq{vardef}. Alternatively, one can assign homogeneous power counting to $\wa$ and $\wb$ using \eq{partonic_expansion}, which then requires to expand the rapidity $Y$ in $\lambda$.
 \item $(p_1^-,p_2^+,k^+,k^-,x)$: Here, we trade $Q^2$ and $Y$ for the lightcone momenta $(p_1^-,p_2^+)$ of the incoming partons. This parameterization has the advantage of expressing everything in terms of the momenta of massless particles, i.e.\ the incoming momenta and the hadronic radiation.
 A disadvantage of this parameterization is that $p_1^-$ and $p_2^+$ do not have manifest power counting in terms of hadronic variables $Q^2$ and $Y$, and thus must be expanded in $\lambda$.
\end{itemize}
These parameterizations are of course equivalent, and in practice the preferred parameterization depends on the intended application. While the general illustration of the power expansion is made most manifest using $(Q^2,Y,k^+,k^-,x)$, expanding the partonic cross sections is simplified using ($Q^2, \wa, \wb, x)$.

Finally, we give the explicit relation between the different parameterizations.
We can change variables from $(\wa,\wb)$ to $(k^+,k^-)$ using
\begin{align} \label{eq:sigma_jacob_2}
 \frac{\df\eta_{ij}}{\df Q^2 \df Y \df k^+ \df k^- \df x} &
 = \frac{z(\wa,\wb)}{Q^2} \frac{\df\eta_{ij}}{ \df Q^2 \df \wa \df \wb \df x}
   \bigg|_{\substack{\wa = \wa(Q,Y,k) \\ \wb = \wb(Q,Y,k)}}
\,,\end{align}
where the required variable transformations are given by
\begin{align}
 \wa(Q,Y,k) &= \frac{-k^-}{p_1^-(Q,Y,k)}
 \,,\quad
 p_1^-(Q,Y,k) = -k^- + e^{+Y}\sqrt{Q^2 + k^+ k^- (1-x)}
\,,\nn\\
 \wb(Q,Y,k) &= \frac{-k^+}{p_2^+(Q,Y,k)}
 \,,\quad
 p_2^+(Q,Y,k) = -k^+ + e^{-Y}\sqrt{Q^2 + k^+ k^- (1-x)}
\,,\nn\\
 z(\wa,\wb) &= 1 - \wa - \wb + x \wa \wb
\,,\end{align}
and for brevity we keep implicit that $k^\mu$ is parameterized in terms of $(k^+,k^-,x)$.
Note that since $k^+$ and $k^-$ are defined in the hadronic center-of-mass frame, manifest boost-invariance is lost, and thus \eq{sigma_jacob_2} becomes explicitly $Y$-dependent.
\Eq{sigma_jacob_2} makes it clear that defining the power counting in terms of $Q^2$ and $k$ requires a expansion of $\wa$ and $\wb$ on the right hand side.

We can further change variables from $(Q^2,Y)$ to $(p_1^-,p_2^+)$,
\begin{align} \label{eq:sigma_jacob_3}
 \frac{\df\eta_{ij}}{\df p_1^- \df p_2^+ \df k^+ \df k^- \df x} &
 = \frac{\df\eta_{ij}}{\df Q^2 \df Y \df k^+ \df k^- \df x}
   \bigg|_{\substack{Q^2 = Q^2(p_1,p_2,k) \\ Y = Y(p_1,p_2,k)}}
\,,\end{align}
where the required variable transformations are given by
\begin{align}
 Q^2 = (p_1^- + k^-) (p_2^+ + k^+) - (1-x) k^+ k^-
\,,\quad
 Y = \frac12 \ln\frac{p_1^- + k^-}{p_2^+ + k^+}
\,.\end{align}
Here, fixing the power counting of $p_1^-$, $p_2^+$ and $k$ requires to expand $Q^2$ and $Y$ accordingly.

%% file: Chapters/ExpansionMatrixElements.tex
%%%%%%%%%%%%%%%%%%%%%%%%%%%%%%%%%%%%%%%%%%%%%%%%%%%%%%%%%%%%%%%%%%%%%%%%%%%%%%%%
\section{Collinear expansion of matrix elements}
\label{sec:collinear_expansion}
%%%%%%%%%%%%%%%%%%%%%%%%%%%%%%%%%%%%%%%%%%%%%%%%%%%%%%%%%%%%%%%%%%%%%%%%%%%%%%%%

In this section we show how the technique of collinear expansions developed in the previous section is applied in practice.
To setup our conventions for this section, we first discuss the phase space volume for producing the hard probe $h$ with additional emissions in \sec{expansions_ps}, before illustrating the collinear expansion of matrix elements explicitly for both real radiation in \sec{expansions_real} and for loop integrals in \sec{expansions_loop}.

Throughout this section, we will consider the scenario where $k^\mu$ is collinear to the incoming parton with momentum $p_1^\mu = p_1^- n^\mu/2$. According to \eq{modes}, this implies that we assign the following scaling to $k$:
\begin{align}
 k^\mu = (k^+,k^-,k_\perp) \sim ( \lambda^2, 1, \lambda)
\,.\end{align}
In order to obtain a strict power series expansion of the hadronic cross section it is necessary to expand the partonic momentum components $p_1^-$ and $p_2^+$ around the collinear limit.
For the purpose of this section we instead perform a \emph{partonic} collinear expansion (see \sec{expandingvariables}), treating $p_{1,2}$ as external variables and thus as $\cO(\lambda^0)$ quantities.
All final results are functions of $k^-/p_1^-$ and $k^+/p_2^+$ and one can straightforwardly recover a pure expansion in terms of hadronic observables following \sec{expandingvariables}.

%===============================================================================
\subsection{Collinear phase space}
\label{sec:expansions_ps}
%===============================================================================

The phase space volume for producing the hard probe $h$ with two emissions, as defined in \eq{Phi_hn}, is given by
\begin{align} \label{eq:PS2}
 \Phi_{h+2}&
 = \int\! \frac{ \df\Phi_{h+2} }{\df \wa\df \wb \df x} 
 = \includegraphics[valign=c,width=4cm]{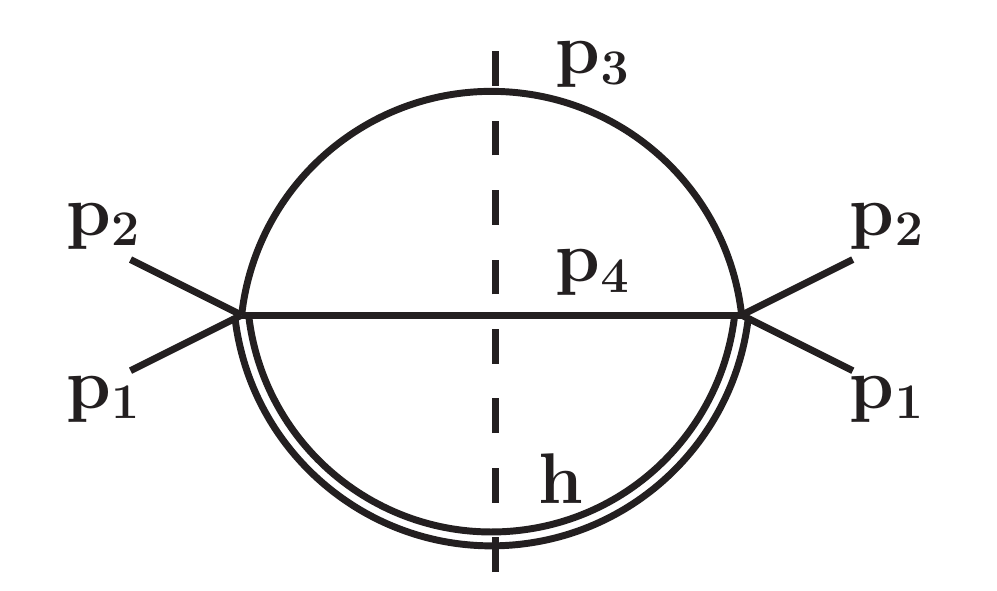} \nn\\
& = \frac{1}{s^2}\frac{ (k^+k^-)^{1-2 \eps } (1-x)^{-\eps } x^{-\eps }}{128 \pi ^3 (1-2 \eps ) \Gamma (1-2 \eps )}
\,.\end{align}
It follows immediately that the phase space volume in \eq{PS2} scales as $\Phi_{h+2} \sim \lambda^{2-4\eps}$.

As usual, we take all momenta to be incoming, and denote the total momentum of all outgoing partons by $k = p_3 + p_4$.
In the above diagram and those below, the dashed line indicates the on-shell constraints of the final state particles, with the solid lines representing massless partons and the double line representing the heavy color-singlet state $h$.

The scaling of  $\Phi_{h+2} \sim \lambda^{2-4\eps}$ can be easily deduced without calculating the actual phase space integrals.
Since $k$ is treated collinear to $p_1$, both final-state momenta $p_3$ and $p_4$ must be collinear to $p_1$ as well.
The associated integration measures and $\delta$ functions entering \eqs{Phi_hn}{phi_2_and_0} transform as
\begin{align}
 \int\df^d p_i \, \delta_+(p_i^2) ~\rightarrow~ \lambda^{2-2\eps} \int\df^d p_i \,\delta_+(p_i^2)
\,,\qquad
 \delta(k^2) ~\rightarrow~ \lambda^{-2} \delta(k^2)
\,.\end{align}
As a consequence, the double-real phase space measure scales as
\begin{align} \label{eq:phi20_coll}
 \int\!\frac{ \df\Phi_{h+2} }{\df \wa\df \wb \df x} ~\rightarrow~ \lambda^{2 - 4\eps} \int\!\frac{ \df\Phi_{h+2} }{\df \wa\df \wb \df x} 
\,,\end{align}
which is precisely the scaling observed in \eq{PS2}.
Similarly, it follows that the more general case of the $h$ + $n$ real emission phase space has the scaling
\begin{align} \label{eq:phin0_coll}
 \int\!\frac{ \df\Phi_{h+n} }{\df \wa\df \wb \df x} ~\rightarrow~ \lambda^{n(2-2\eps) - 2}  \int\!\frac{ \df\Phi_{h+n} }{\df \wa\df \wb \df x} 
\,.\end{align}

%===============================================================================
\subsection{Collinear limit of real radiation}
\label{sec:expansions_real}
%===============================================================================

We consider the following example of a more complicated, purely real Feynman integral,
\begin{align} \label{eq:RR}
 I_{\rm RR}
 = \includegraphics[valign=c,width=4cm]{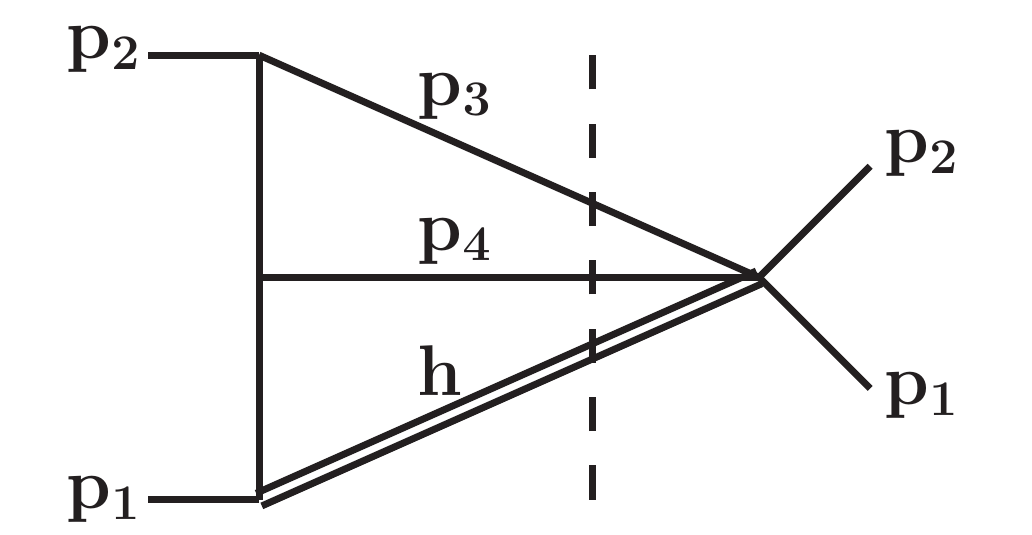}
 = \int\frac{ \df\Phi_{h+2} }{\df \wa\df \wb \df x}  \frac{1}{(p_2+p_3)^2(p_2+p_3+p_4)^2}
\,.\end{align}
Let us first consider the case where both $p_3$ and $p_4$ are collinear to $p_2$.
In this scenario, since  both propagators in \eq{RR} only involve collinear momenta, and thus scale homogeneously as $\lambda^2$ under the $\bn$-collinear rescaling of \eq{modes} and no expansion of \eq{RR} in $\lambda$ is needed.

In contrast, if we consider $p_3$ and $p_4$ to be collinear to $p_1$, then the second propagator is not homogeneous in $\lambda$ anymore, as it contains both $n$-collinear and $\bn$-collinear momenta.
To expand the propagator in this limit, we apply the $n$-collinear rescaling of \eq{modes} to $p_3$ and $p_4$,
\begin{align} \label{eq:p34_nbcoll}
 p_{3,4}^\mu ~\to~  p_{3,4}^- \frac{n^\mu}{2} + \lambda^2\,p_{3,4}^+ \frac{\bn^\mu}{2} + \lambda\,\, p_{3,4\,\perp}^\mu
\,.\end{align}
With these rescalings, it is now straightforward to expand the second propagator in \eq{RR} in $\lambda$,
\begin{align} \label{eq:RR_expansion}
 \frac{1}{(p_2+p_3+p_4)^2} &
 = \frac{1}{p_2^+ (p_3^- + p_4^-) + 2 p_3 \cdot p_4}
 ~\stackrel{p_1-\rm coll}{\longrightarrow}~
 \frac{1}{p_2^+ (p_3^- + p_4^-) + \lambda^2 \, 2 p_3 \cdot p_4}
\nn\\&
 = \sum_{n=0}^\infty (\lambda^2)^n \frac{(-2 p_3 \cdot p_4)^n}{\bigl[p_2^+ (p_3^- + p_4^-) \bigr]^{n+1}}
\,.\end{align}
For $n=0$, this propagator is linear in the real momenta $p_3$ and $p_4$, and thus corresponds to an eikonal propagator.
Higher orders in $\lambda$ only involve pure powers of the eikonal propagator, thus yielding a relatively simple structure of the expansion.
Together with \eq{phi20_coll}, the integral in \eq{RR} can thus be expanded as
\begin{align} \label{eq:RR2}
 I_{\rm RR} &
 ~\stackrel{p_1-\rm coll}{\longrightarrow}~
 \sum_{n=0}^\infty (\lambda^2)^{n+1-2\eps}
   \int\frac{ \df\Phi_{h+2} }{\df \wa\df \wb \df x} \frac{(-2 p_3 \cdot p_4)^i}{(p_2+p_3)^2 \bigl[p_2^+ (p_3^- + p_4^-) \bigr]^{n+1}}
\,.\end{align}
This expansion can be represented diagrammatically as
\begin{align}
 \includegraphics[valign=c,width=3cm]{Diagrams/RR.pdf} ~\rightarrow~
 \lambda^{2-4\eps} \Biggl[ \includegraphics[valign=c,width=3cm]{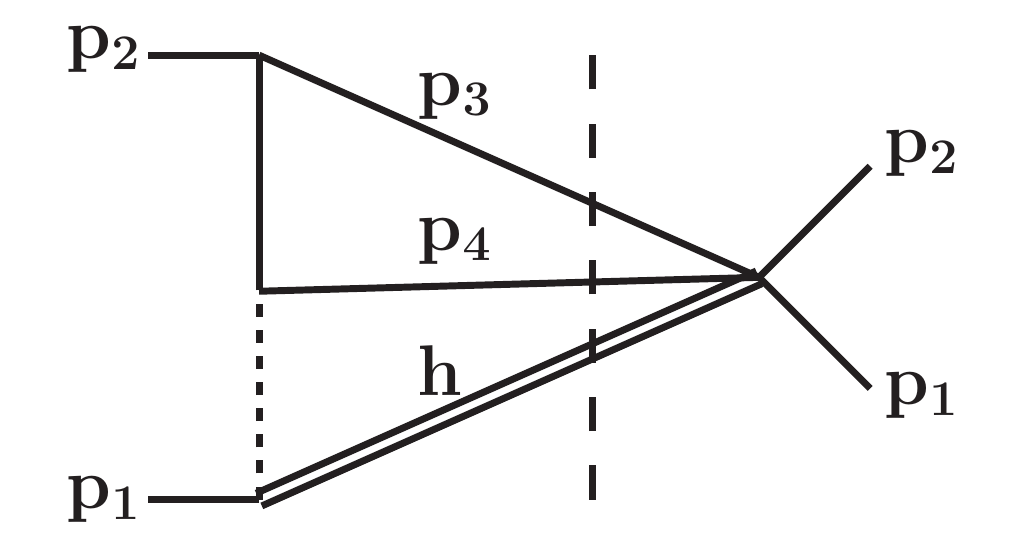}
   - \lambda^2 \includegraphics[valign=c,width=3cm]{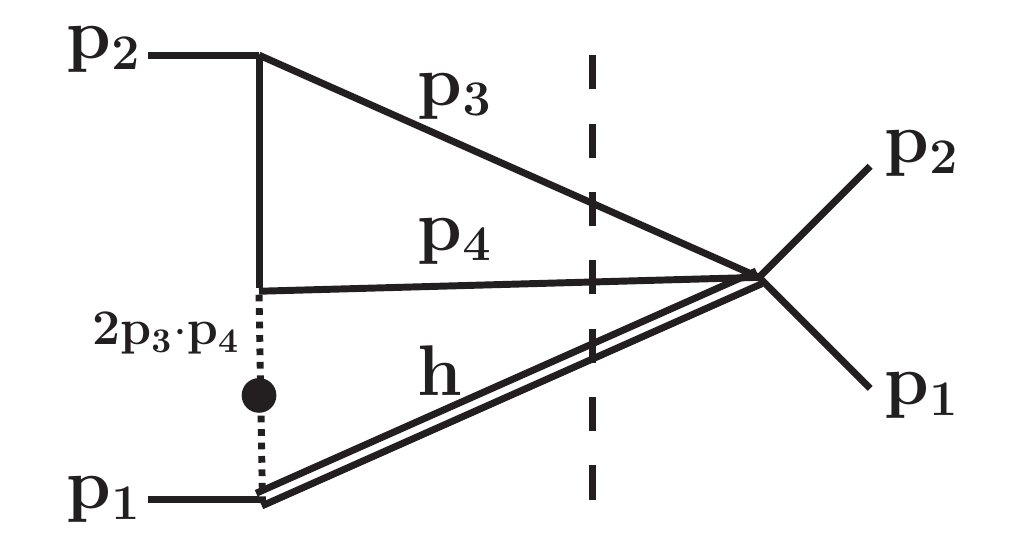}
   + \cO(\lambda^3)
 \Biggr]
\,.\end{align}
Here, the dotted line indicates the expanded (eikonal) propagator and the dot on the line represents higher powers of this propagator. The label denotes the additional kinematic factor arising from \eq{RR_expansion}.

The expansion in \eq{RR2} results in several advantages.
First, we observe that each term in the expansion is homogeneous under the $n$-collinear rescaling transformation in \eq{p34_nbcoll}.
As a consequence, we may directly determine the functional dependence of each term in the expansion on $k^+$ similarly to the case of the phase space volume.
In other words, the resulting functions will be simpler since they only depend on $k^+$ via a multiplicative pre-factor.
Second, the structure of expanded Feynman integrals is amenable to IBP reduction techniques via the framework of reverse unitarity~\cite{Anastasiou2003,Anastasiou:2002qz,Anastasiou:2003yy,Anastasiou2005,Anastasiou2004a}.
The benefit is that the appearing integrals can be related to so called master integrals. 
In our example we find the IBP relations
\begin{alignat}{3} \label{eq:coefeqs}
 \includegraphics[valign=c, width=4cm]{Diagrams/RRColl0.pdf} &= -\frac{1-2\eps}{\eps (p_2^+ k^-)^2 } &&\times  \includegraphics[valign=c, width=4cm]{Diagrams/RRPS.pdf}
%%%
\,,\\ \label{eq:coefeqs2}
%%%
 \includegraphics[valign=c, width=4cm]{Diagrams/RRColl1.pdf} &= -\frac{k^+ x}{p_2^+}\frac{1-2\eps}{\eps(p_2^+ k^-)^2 }  &&\times  \includegraphics[valign=c, width=4cm]{Diagrams/RRPS.pdf}
\,.\end{alignat}
Clearly, it is very advantageous that any higher order term in our expansion is related to the same master integrals as the first, which in our example is just the phase space volume.
The unexpanded integral of our example in \eq{RR} is itself related to the phase space volume by an IBP identity,
\begin{align}
 \includegraphics[valign=c, width=4cm]{Diagrams/RR.pdf} &= -\frac{(1-2 \eps) }{\eps (p_2^+ k^-)^2 }\left(1+\frac{k^+ x}{p_2^+}\right)^{-1}  \times  \includegraphics[valign=c, width=4cm]{Diagrams/RRPS.pdf}
\,.\end{align}
From this we can easily see that the coefficients obtained in \eq{coefeqs} and \eq{coefeqs2} correspond exactly to the coefficients of the expansion of the exact result.

In summary, we outlined a procedure that allows us to perform an expansion of real radiation integrals around the limit of all final state partons becoming collinear to an initial state momentum.
This expansion is carried out by simply performing the appropriate collinear rescaling transformation of \eq{modes} on all final state parton momenta and subsequently expanding the integrand of our real radiation integral in the artificial parameter $\lambda$, prior to actually evaluating the integral.
Each term in the expansion in $\lambda$ then corresponds to exactly one term in the expansion of the integral in $k^+$. 
The computation of the terms in the expansion is greatly facilitated by applying techniques like IBP identities via the reverse unitarity framework.
In particular, any term appearing at higher orders in the expansion will be expressible in terms of master integrals that appear already in the first few terms.

%===============================================================================
\subsection{Expansion of loop integrals}
\label{sec:expansions_loop}
%===============================================================================

In contrast to the phase space integral over real momenta considered in \sec{expansions_real}, where the requirement of $k$ being collinear to $p_1$ restricted $p_{3,4}$ to be collinear to $p_1$ as well, such a restriction does not appear for loop momenta.
Despite this, it is still useful to expand loop integrals in a similar fashion around the hard, collinear and soft regions.
As discussed in \sec{expansions_intro}, for factorization proofs this is crucial to separate these different regions into distinct matrix element,
while in the method-of-regions approach of \refcite{Beneke:1997zp} it used to simplify loop integrals by expanding the integrand in all relevant limits and combining their individual results.

Here, we will show for a simple example how one can easily approximate and expand loop integrals in the discussed regimes, and that the sum of all regions indeed reproduces the full result. This will be illustrated using the following real-virtual diagram,
\begin{align} \label{eq:RV}
 I_{\rm RV} &
 = \includegraphics[valign=c,width=4cm]{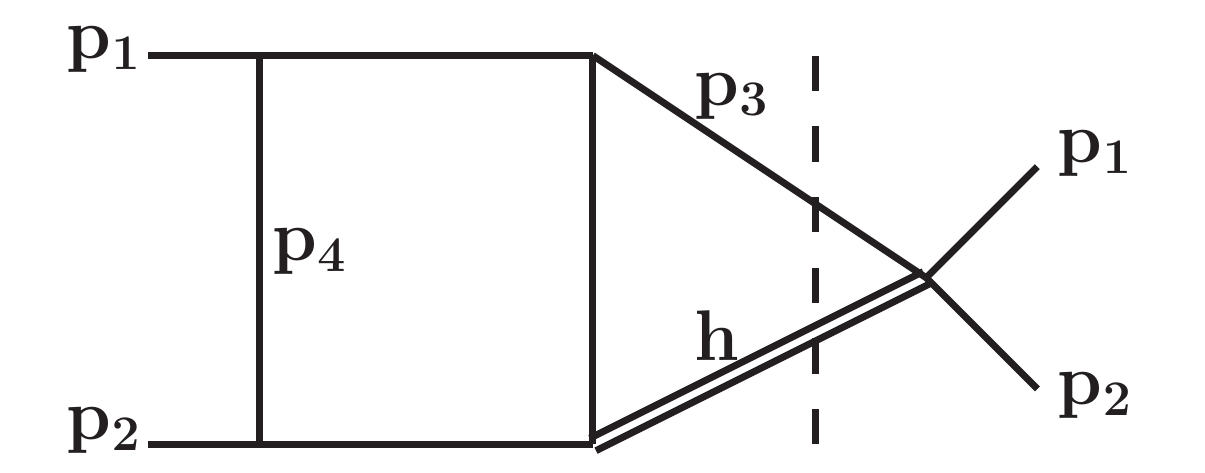}
 = \int\frac{ \df\Phi_{h+1} }{\df \wa\df \wb \df x} \frac{\df^d p_4}{(2\pi)^d}
   \frac{1}{p_4^2 \, (p_1 + p_4)^2 \, (p_1 + p_3 + p_4)^2 \, (p_2 - p_4)^2}
\nn\\&
 = \frac{\img c_\Gamma}{128 \pi^4 \eps^2}\delta(x) \biggl[
   \frac{(k^+ k^-)^{-\eps}}{s^4 (-s)^{\eps}} \sum_{n,m=0}^\infty\frac{(\eps+1)_n \, (\eps+2)_{m+n}}{(n+m+1) \, m!\,n!\, (\eps+2)_n}  \left(\frac{k^+}{p_2^+}\right)^{m} \left(\frac{k^-}{p_1^-}\right)^{n}
 \nn\\&\hspace{2.7cm}
 - \frac{(k^+ k^-)^{-\eps}}{s^3 (p_1^- k^+)^{1+\eps}} (4 \pi)^{-2\eps}\, _2F_1\left(1,-\eps; 1-\eps; \frac{k^-}{p_1^-}\right)
   \biggr]
\,.\end{align}
Here, the total final-state hadronic momentum is $k = p_3$, and thus the $p_3$ integral is actually fixed.
In \eq{RV}, $(a)_n = \Gamma(a+n)/\Gamma(a)$ is the (rising) Pochhammer symbol, and we abbreviate common loop factors by
\begin{align}
 c_\Gamma  = \frac{\Gamma(1+\eps) \Gamma(1-\eps)}{\Gamma(1-2\eps)}
\,.\end{align}
As before, we consider the limit where $k$ is collinear to $p_1$, such that $k^+ \sim \cO(\lambda^2)$ and $k^- \sim \cO(\lambda^0)$.
This immediately implies that \eq{RV} scales as
\begin{align} \label{eq:RV_expanded}
 I_{\rm RV} &\stackrel{\rm coll}{\longrightarrow}
\delta(x)  \frac{\img c_\Gamma }{128 \pi^4 \eps^2} \biggl[
   \lambda^{-2\eps} \frac{(k^+ k^-)^{-\eps}}{s^4 (-s)^{\eps}} \sum_{n,m=0}^\infty \lambda^{2m} \frac{(\eps +1)_n \, (\eps +2)_{m+n}}{(n+m+1) \, m!\,n!\, (\eps +2)_n}  \left(\frac{k^+}{p_2^+}\right)^{m} \left(\frac{k^-}{p_1^-}\right)^{n}
 \nn\\&\hspace{3cm}
 - \lambda^{-2-4\eps} \frac{(k^+ k^-)^{-\eps}}{s^3 (p_1^- k^+)^{1+\eps}} (4 \pi)^{-2\eps}\, _2F_1\left(1,-\eps; 1-\eps;  \frac{k^-}{p_1^-}\right)
   \biggr]
\,.\end{align}
The second line has homogeneous scaling in $\lambda^{-2-4\eps}$, and is the dominant contribution in the limit $\lambda\to0$.
We will see below that this result is entirely from the region where the loop momentum is collinear to $p_1$.
In other words, the leading-power limit of $I_{\rm RV}$ arises from the region where both loop \emph{and} real momenta are collinear to $p_1$.
The first line in \eq{RV_expanded} does not scale homogeneously in $\lambda$, but is suppressed at least as $\cO(\lambda^2)$ compared to the leading-power limit. We will see that this line entirely arises from the region where the loop momentum is hard.
In particular, the two contributions have different fractional scalings in $\lambda^{-2\eps}$ and $\lambda^{-4\eps}$, respectively.
These scalings arise entirely from the loop integral measures, and thus can be easily distinguished between the different contributions.

%===============================================================================
\subsubsection{Collinear limit}
%===============================================================================

We first consider the loop momentum $p_4$ to be collinear to the incoming parton with momentum $p_1$.
According to \eq{modes}, we hence transform
\begin{align}
 p_4^\mu ~\to~  p_4^- \frac{n^\mu}{2} + \lambda^2\, p_4^+ \frac{\bn^\mu}{2} + \lambda\,\, p_{4\perp}^\mu
\,.\end{align}
The first three propagators in \eq{RV} scale homogeneously as $\cO(\lambda^{-2})$ under this rescaling,
while the last propagator in \eq{RV} is not homogeneous in $\lambda$ and must be expanded,
\begin{align} \label{eq:coll_prop_exp}
 \frac{1}{(p_2 - p_4)^2}  &
 = \frac{1}{-2 p_2 \cdot p_4 + p_4^2}
 ~\stackrel{p_1-\rm coll}{\longrightarrow}~
 \frac{1}{-2 p_2 \cdot p_4 + \lambda^2 p_4^2}
 = \sum_{n=0}^\infty \lambda^{2n} \frac{(-p_4^2)^n}{(-2 p_2 \cdot p_4)^{n+1}}
\,.\end{align}
Together with \eq{phin0_coll}, this allows us to expand the integrand in \eq{RV} as
\begin{align} \label{eq:RV_ncoll}
 I_{\rm RV}
 \stackrel{p_1-\rm coll}{\longrightarrow}&~
 \lambda^{-2-4\eps} \sum_{n=0}^\infty \lambda^{2n} \!\!\int\!\!\df\Phi_{h+1} \frac{\df^d p_4}{(2\pi)^d} \frac{(-p_4^2)^n}{p_4^2 \, (p_1 + p_4)^2 \, (p_1 + p_3 + p_4)^2 \, (-2 p_2 \cdot p_4)^{n+1}}
 \\\nn=&~
 \lambda^{-2-4\eps} \biggl[~
   \int\frac{ \df\tilde\Phi_{h+1} }{\df \wa\df \wb \df x}  \frac{\df^d p_4}{(2\pi)^d} \frac{1}{p_4^2 \, (p_1 + p_4)^2 \, (p_1 + p_3 + p_4)^2 \, (-2 p_2 \cdot p_4)}
   \\\nn&\hspace{1.2cm}
   - \lambda^2 \int\frac{ \df\Phi_{h+1} }{\df \wa\df \wb \df x}  \frac{\df^d p_4}{(2\pi)^d} \frac{1}{(p_1 + p_4)^2 \, (p_1 + p_3 + p_4)^2 \, (-2 p_2 \cdot p_4)^2} + \cO(\lambda^4) \biggr]
\,.\end{align}
The overall scaling in $\lambda^{-4\eps}$ arises from $\df^d p_4 \sim \lambda^{4-2\eps}$ and $\df\Phi_{h+1} \sim \lambda^{-2\eps}$, and thus is independent of the structure of the diagram itself.
Each order of the expanded integrand is now homogeneous in $\lambda$.
The expansion in \eq{RV_ncoll} can be illustrated graphically as
\begin{align} \label{eq:RV_ncoll_diag}
 I_{\rm RV} &
 \stackrel{\rm coll}{\longrightarrow}
 \lambda^{-2-4\eps} \biggl[ \includegraphics[valign=c,width=4cm]{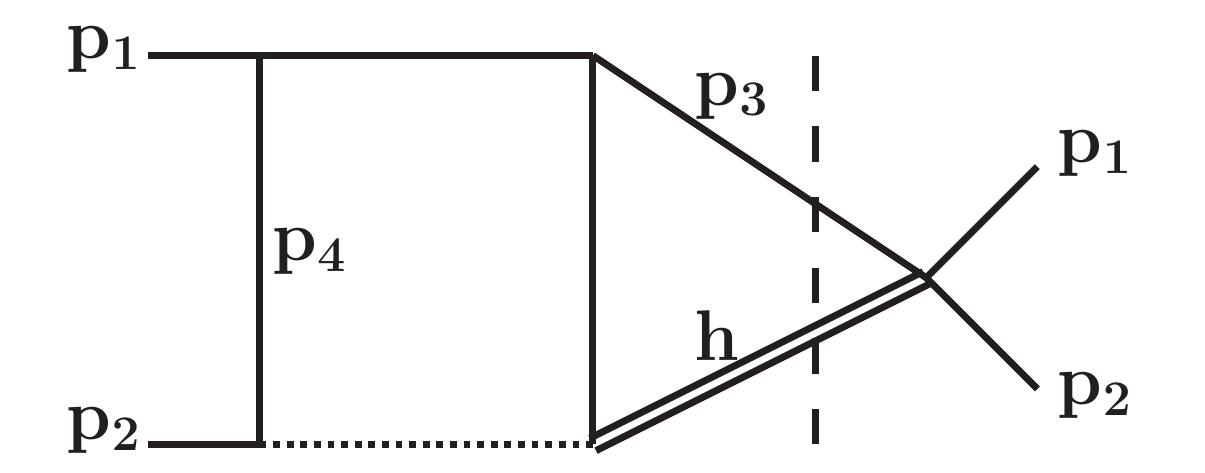}
 - \lambda^2 \includegraphics[valign=c,width=4cm]{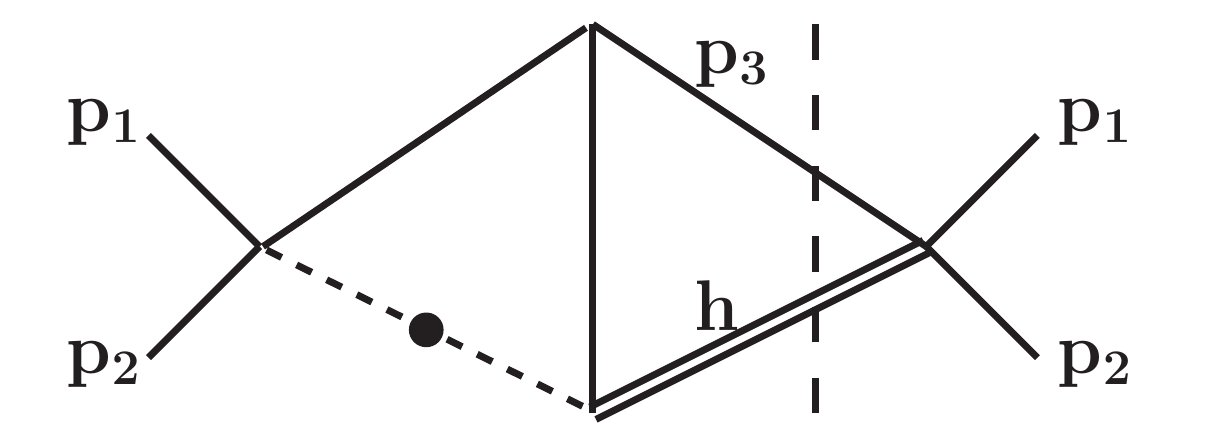} + \cO(\lambda^4) \biggr]
\,.\end{align}
Here, the dotted lines are linear (eikonal) propagators, and the dot on the line denotes that the propagator is raised to one power.
Note that in the second diagram, the explicit $1/p_4^2$ propagator is canceled, indicated by the contracted vertex.
Being able to represent collinear expanded diagrams again in a diagrammatic fashion is extremely useful.
In particular, the structure observed in the collinear loop expansion makes it possible to use IBP techniques for the computation of loop and phase space integrals.

The leading-power integral in \eq{RV_ncoll} can be evaluated as
\begin{align} \label{eq:RV_ncoll_LP}
 I_{\rm RV} & \stackrel{\rm coll}{=}
 - \delta(x) \lambda^{-2-4\eps} \frac{\img c_\Gamma}{128 \pi^4 \eps^2} \frac{(k^+ k^-)^{-\eps}}{s^3 (p_1^- k^+)^{1+\eps}}  (4 \pi)^{-2\eps}  {}_2F_1\left(1,-\eps; 1-\eps; \frac{k^-}{p_1^-}\right) \times \bigl[1 + \cO(\lambda^2) \bigr]
\,,\end{align}
and thus correctly reproduces the last line of \eq{RV_expanded}.
Note that the higher-order terms in $\lambda$, such as the second integral in \eq{RV_ncoll}, can be shown to vanish identically in dimensional regularization.

%===============================================================================
\subsubsection{Hard limit}
%===============================================================================

The hard region is characterised by treating the loop momentum as uniformly larger than our expansion parameter $\lambda$, while the final state momentum $p_3$ is still treated as collinear to $p_1$.
Only one propagator in \eq{RV} involves $p_3$, and can be expanded in $\lambda$ as
\begin{align} \label{eq:hardpropexp}
 \frac{1}{(p_1 + p_3 + p_4)^2} =&~
 \frac{1}{\bigl[  p_4^2 + p_4^+(p_1^- + p_3^-)  \bigr] + p_3^+ (p_1^- + p_4^-) + 2 p_{3\perp} \cdot p_{4\perp}}
 \nn\\ \underset{p_3 \, \parallel \, p_1}{\overset{p_4\, \rm hard}{\longrightarrow}}&~
 \frac{1}{\bigl[  p_4^2 + p_4^+(p_1^- + p_3^-)  \bigr] + \lambda^2 p_3^+ (p_1^- + p_4^-) + 2 p_{3\perp} \cdot p_{4\perp}}
 \nn\\ =&~
 \sum_{n=0}^\infty (-\lambda)^n \frac{\bigl[\lambda p_3^+ (p_1^- + p_4^-) + 2 p_{3\perp} \cdot p_{4\perp} \bigr]^n}{\bigl[ p_4^+ (p_1^- + p_3^-) + p_4^2 \bigr]^{n+1}}
\,.\end{align}
All other propagators in \eq{RV} scale as $\cO(\lambda^0)$ and are not expanded.
Together with the rescaling of the phase space measure according to \eq{phin0_coll}, the leading-power hard limit of \eq{RV} becomes
\begin{align} \label{eq:RV_hard}
 I_{\rm RV} &
 ~\stackrel{\rm hard}{\longrightarrow}~
 \lambda^{-2\eps} \int\frac{ \df\Phi_{h+1} }{\df \wa\df \wb \df x}  \frac{\df^d p_4}{(2\pi)^d} \frac{1}{p_4^2 \, (p_1 + p_4)^2 \, (p_2 - p_4)^2 \bigl[ p_4^+ (p_1^- + p_3^-) + p_4^2 \bigr]}
 \times \bigl[1 + \cO(\lambda) \bigr]
\,.\end{align}
The overall scaling in $\lambda^{-2\eps}$ arises entirely from the phase space measure, as the hard loop measure scales as $\df^d p_4 \sim \lambda^0$. This shows that hard and collinear loops never have the same dependence on $\eps$, and thus can be easily distinguished by their overall scalings.

Despite the modified propagator in this integral, it can still be subjected to the usual loop integration techniques like IBPs and differential equations.
The same holds true for all higher order terms in the expansion of the full loop integral.
The explicit example in \eq{RV_hard} is also easily performed using Feynman parameters. We obtain
\begin{align} \label{eq:RV_hard2}
 I_{\rm RV} &~\stackrel{\rm hard}{\longrightarrow}~
 \frac{\img c_\Gamma}{128 \pi^4 \eps (1+\eps)} \delta(x)  \lambda^{-2\eps}\frac{(k^+k^-)^{-\eps }}{p_2^+ k^- s^{3+\eps}}
 \left[1-\left(1+\frac{k^-}{p_1^-}\right)^{-1-\eps}\right]
 \times \bigl[1 + \cO(\lambda) \bigr]
\,.\end{align}
This result exactly agrees with the infinite sum over $n$ in \eq{RV_expanded} evaluated for $m=0$, \emph{i.e.}\ the $\cO(\lambda^{-2\eps})$ to \eq{RV_expanded}.
Furthermore, every higher-order term in the expansion in $k^+$ of the second to last line of \eq{RV} corresponds to exactly one term in the integrand expansion of $I_{\rm RV}$ in $\lambda$.
Terms proportional to odd powers of $\lambda$ drop out identically.
Since higher order terms in the expansion essential just modify the powers of the propagators at the integrand level according to \eq{hardpropexp} it is particularly convenient to use IBP techniques in such a computation.

%===============================================================================
\subsubsection{Anticollinear limit}
%===============================================================================

For completeness, we also consider the limit where $p_4$ is collinear to the incoming parton with momentum $p_2$,
in contrast to $k$ which is chosen collinear to $p_1$.
According to \eq{modes}, we hence transform
\begin{align}
 p_4^\mu ~\to~ \lambda^2\, p_4^- \frac{n^\mu}{2} + p_4^+ \frac{\bn^\mu}{2} + \lambda\,\, p_{4\perp}^\mu
\,,\qquad
 \df^d p_4 ~\to~ \lambda^d \df^d p_4
\,.\end{align}
With this rescaling, we need to expand two propagators of the integrand in \eq{RV},
\begin{align} \label{eq:bncoll_prop_exp}
 \frac{1}{(p_1 + p_4)^2} &\stackrel{p_2-\rm coll}{\longrightarrow} \frac{1}{p_1^- p_4^+} \times \bigl[1 + \cO(\lambda^2)\bigr]
\,,\nn\\
 \frac{1}{(p_1 + p_3 + p_4)^2} &\stackrel{p_2-\rm coll}{\longrightarrow} \frac{1}{(p_1^- + p_3^-) p_4^+} \times \bigl[1 + \cO(\lambda^2)\bigr]
\,,\end{align}
For brevity, we only show the two leading terms each.
Together with \eq{phin0_coll}, this allows us to expand the integrand in \eq{RV} as
\begin{align} \label{eq:RV_nbcoll}
 I_{\rm RV} &
 \stackrel{p_2-\rm coll}{\longrightarrow}
 \lambda^{-4\eps} \int\frac{ \df\Phi_{h+1} }{\df \wa\df \wb \df x}  \frac{\df^d p_4}{(2\pi)^d}
   \frac{1}{p_4^2 \, (p_1^- p_4^+) \, [(p_1^- + p_3^-) p_4^+] \, (p_2 - p_4)^2} \times \bigl[1 + \cO(\lambda^2) \bigr]
\,,\end{align}
which is scaleless and thus vanishes in pure dimensional regularization.
Note that the integral from expanding the propagators through $\cO(\lambda^n)$ scales as $\lambda^{n-4\eps}$.
Since the only term with this $\eps$ dependence in \eq{RV_expanded} is fully given by the $n$-collinear limit of \eq{RV_ncoll_LP}, the $p_2$-collinear limit must in fact vanish to all orders in $\lambda$.

%===============================================================================
\subsubsection{Soft limit}
\label{sec:soft_limit}
%===============================================================================
We want to compare the result of the collinear expansion to a soft expansion of Feynman diagrams.
To obtain the purely soft region, we rescale the loop momentum $p_4$ in \eq{RV} as
\begin{align} \label{eq:p4_soft}
 p_4^\mu
 ~\stackrel{\rm soft}{\longrightarrow}~ \lambda^2 p_4^- \frac{n^\mu}{2} + \lambda^2 p_4^+ \frac{\bn^\mu}{2} + \lambda^2 p_{4\perp}^\mu
\,.\end{align}
To obtain the soft-collinear overlap, we first rescale $p_4$ as collinear, followed by a subsequent soft rescaling,
\begin{align} \label{eq:p4_zerobin}
 p_4^\mu
 ~\stackrel{\rm coll}{\longrightarrow}~ p_4^- \frac{n^\mu}{2} + \lambda^2 p_4^+ \frac{\bn^\mu}{2} + \lambda p_{4\perp}^\mu
 ~\stackrel{\rm soft}{\longrightarrow}~ \lambda^2 p_4^- \frac{n^\mu}{2} + \lambda^2 p_4^+ \frac{\bn^\mu}{2} + \lambda^2 p_{4\perp}^\mu
\,.\end{align}
Let us explicitly discuss the transformation of two of the propagators in \eq{RV} under \eq{p4_zerobin},
\begin{alignat}{3} \label{eq:blubb_1}
 &\frac{1}{(p_2 - p_4)^2} = \frac{1}{p_4^2 - p_2^+ p_4^-}
 && ~\stackrel{\rm coll}{\longrightarrow}~ \frac{1}{\lambda^2 p_4^2 - p_2^+ p_4^-}
 && ~\stackrel{\rm soft}{\longrightarrow}~ \frac{1}{\lambda^2 (\lambda^2\, p_4^2 - p_2^+ p_4^-)}
%%%
\,,\nn\\
%%%
 &\frac{1}{(p_1 + p_4)^2} = \frac{1}{p_4^2 + p_1^- p_4^+}
 && ~\stackrel{\rm coll}{\longrightarrow}~ \frac{1}{\lambda^2 (p_4^2 + p_1^- p_4^+)}
 && ~\stackrel{\rm soft}{\longrightarrow}~ \frac{1}{\lambda^2 (\lambda^2 \, p_4^2 + p_1^- p_4^+)}
\,.\end{alignat}
In the first case, rescaling the collinear limit as soft only amounts to an overall rescaling by $\lambda^{-2}$, but does not change the relative scaling of the two terms in the propagator.
In the second case, we observe both that only one term in the denominator gets rescaled in the soft limit, and thus  one will encounter a different kinematic structure when expanding this propagator in $\lambda$ than in the collinear limit.
However, in both cases shown in \eq{blubb_1}, it is easy to see that the soft-collinear limit is identical to taking the soft limit directly.
The same holds for the two other propagators in \eq{RV} that are not explicitly shown here.
In conclusion, we find that at the diagram level, the soft-collinear overlap is equal to soft limit itself.

Finally, we note that the leading-power soft limit of \eq{RV} is given by
\begin{align} \label{eq:RV_soft}
 I_{\rm RV} &~\stackrel{\rm soft}{\longrightarrow}~
 \lambda^{-2-6\eps} \int\frac{ \df\tilde\Phi_{h+1} }{\df \wa\df \wb \df x}  \frac{\df^d p_4}{(2\pi)^d}
   \frac{1}{p_4^2 \, (p_1^- p_4^+) \, [(p_1^- + p_3^-) p_4^+ + p_1^- p_3^+ ] \, (-p_2^+ p_4^-)}
   \times \bigl[1 + \cO(\lambda) \bigr]
\,.\end{align}
This integral is scaleless and vanishes in dimensional regularization.

%===============================================================================
\subsection{Discussion}
\label{sec:expansions_discussion}
%===============================================================================

To summarize the key results of this section, we have shown how Feynman diagrams can be systematically expanded in their collinear limit by assigning the appropriate scalings to all loop and real momenta, which allows one to expand the integrand in $\lambda$.
In particular, the expanded integrands allow for a diagrammatic representation and are amenable to standard integral techniques such as IBP~\cite{Chetyrkin:1981qh,Tkachov:1981wb} relations or the method of differential equations~\cite{Kotikov:1990kg,Kotikov:1991hm,Kotikov:1991pm,Henn:2013pwa,Gehrmann:1999as}.
This significantly simplifies evaluating the expanded integrals compared to the exact integral, and thus provides a convenient strategy to approximate Feynman diagrams in the collinear limit.
The illustrated methods are conceptually very simple, and thus easily extend to more complicated diagrams with additional external partons or multi-loop integrals.

In the case of real radiation, the requirement that the total real momentum $k^\mu$ is collinear implies that all real momenta are collinear individually.
This does not apply for loop momenta, which are not confined to be in the collinear region. 
As a consequence we need to follow the method of regions~\cite{Beneke:1997zp} and compute the regions where the loop momenta are hard and where they are collinear.
The sum of both regions yields the correct expansion of our Feynman integrals.
The results of the different regions give rise to different scalings as $\lambda^{-4\eps}$ and $\lambda^{-2\eps}$, respectively. 
This difference is entirely due to the loop measure, and thus hard and collinear contributions can be easily identified by their scaling exponent.
In other words, since the expansion of the loop integrand itself is a simple Laurent series in $\lambda$, the loop measure fully determines the non-integer powers of $\lambda^{-n \eps}$.

We also discussed the soft limit of matrix elements. 
We found in an explicit example that the soft region of a loop integral can be obtained by first computing the collinear region of this integral and subsequently taking the soft limit. 
As a matter of fact this property holds more generally. 
The soft-collinear overlap of a partonic coefficient function can either be computed by first performing the collinear expansion and then the soft expansion, or vice versa.
More precisely, expanding the first $n$ terms in the collinear expansion around the production threshold up to $n$ terms will correctly reproduce the $n^{\text{th}}$ power in the threshold expansion.
This provides a stringent test of the collinear expansion by comparing to existing analytic results for which a threshold expansion was performed.
It also provides a considerable simplification for the calculation of collinear master integrals.
For example, if the method of differential equations is utilized to compute master integrals for the collinear expansion, then the boundary conditions for these differential equations can be chosen to be the threshold-expanded integrals.
For the computation of threshold expanded integrals see for example \refscite{Anastasiou:2013srw,Anastasiou:2015yha,Zhu:2014fma}.

%% file: Chapters/SCET.tex
\section{Kinematic expansions and SCET}
\label{sec:SCET}

Differential cross sections may be kinematically enhanced in all different momentum regions shown in \eq{modes}. 
Above we only discussed the expansion cross sections around one particular limit, namely the collinear limit.
However, in order to perform a physically sensible and consistent expansion of a hadronic cross section we need to expand in observable quantities.
A collinear expansion of a hadronic cross section alone typically does not satisfy this requirement.
In order to obtain a physical expansion in an observable all momentum regions where the observable is kinematically enhanced must be considered.
Depending on the observable of interest, the necessary ingredients to achieve this goal may vary. 

Soft-Collinear Effective Theory (SCET)~\cite{Bauer:2000ew, Bauer:2000yr, Bauer:2001ct, Bauer:2001yt, Bauer:2002nz} provides an excellent tool to organize the expansion in such kinematic limits, and we discuss in \sec{facthm} how the tools developed in the previous section connect to factorization theorems derived in SCET.
In such factorization theorems, it is crucial to account for the overlap of regions when combining multiple kinematic expansions, which we address in \sec{zero_bin}.

%===============================================================================
\subsection{Kinematic expansions and factorisation theorems}
\label{sec:facthm}
%===============================================================================

The momentum regions shown in \eq{modes} are precisely the basis for the formulation of SCET, which is an effective field theory describing QCD in its collinear and soft limits, \emph{i.e.}~the leading infrared region.
Schematically, the SCET Lagrangian is expanded as
\begin{align} \label{eq:LSCET}
 \cL_{\rm SCET} &= \cL_{\rm SCET}^{(0)} + \sum_{k>0} \cL^{(k)}
\,.\end{align}
Here, the superscript $^{(0)}$ indicates the leading-power (LP) terms in the expansion in $\lambda \ll 1$,
where as before $\lambda$ is an auxiliary power counting parameter.
The $\cL^{(k)}$ indicate subleading power Lagrangians~\cite{Manohar:2002fd,Beneke:2002ph,Pirjol:2002km,Beneke:2002ni,Bauer:2003mga,Feige:2017zci,Moult:2017rpl,Chang:2017atu,Moult:2017xpp,Beneke:2017ztn,Beneke:2018rbh} that are suppressed by $\lambda^k$ w.r.t.\ to the leading power.
The leading-power SCET Lagrangian can be organized as
\begin{align} \label{eq:LSCET_LP}
 \cL_{\rm SCET}^{(0)} &= \cL_h^{(0)} + \cL_n^{(0)}  + \cL_\bn^{(0)} + \cL_s^{(0)} + \cL_\cG^{(0)}
\,.\end{align}
Here, $\cL^{(0)}_h$ contains the hard scattering operators mediating the underlying hard interaction,
and $\cL^{(0)}_{n,\bn,s}$ are the SCET Lagrangians for $n$-collinear, $\bn$-collinear and soft fields as defined by \eq{modes}, respectively.%
\footnote{For soft modes, $m=1$, this is referred to as SCET$_{\rm II}$~\cite{Bauer:2002aj}, otherwise as SCET$_{\rm I}$.}
More generally, in the presence of multiple collinear directions as required e.g.\ for multijet processes, \eq{LSCET_LP} contains a sum over all relevant collinear directions $\{n_i\}$.
SCET also allows for a treatment of Glauber modes, which appear as non-local potentials in $\cL_\cG^{(0)}$, the leading power Glauber Lagrangian~\cite{Rothstein:2016bsq}.

In SCET, factorization is achieved by a field redefinition of soft and collinear fields which decouples the soft and collinear Lagrangians from each other \cite{Bauer:2001yt}. These modes can still interact with each other through the Glauber Lagrangian $\cL_\cG^{(0)}$, which thus can break factorization. In this work we will consider observables where the Glauber contributions from $\cL_\cG^{(0)}$ either cancel identically \cite{Bodwin:1984hc,Collins:1984kg,Collins:1985ue,Collins:1988ig,Collins:1989gx,Collins:1350496,Diehl:2015bca} or start contributing to higher perturbative orders that the one we consider in this work \cite{Gaunt:2014ska,Zeng:2015iba}.

In SCET, the leading kinematic regions are made manifest and decoupled from each other at the Lagrangian level, which greatly simplifies the derivation of factorization formulas.
For suitable factorizable infrared-sensitive observables $\Obs$, which we take to vanish as $\Obs\to0$ in the Born limit, the hadronic cross section \eq{sigma_hadr_finite} can be factorized as \cite{Collins:1984kg, Stewart:2009yx}
\begin{equation}  \label{eq:fact_thm}
\frac{\df\sigma}{\df Q^2 \df Y \df \Obs}
= \sigma_0 \sum_{i,j} H_{ij}(Q^2)\, \bigl[B_i(x_1^B,\Obs)\otimes B_j(x_2^B,\Obs)\otimes S(\Obs) \bigr]
\times \bigl[1 + \cO(\Obs/Q) \bigr]
\,.\end{equation}
As usual, $Q$ and $Y$ are the invariant mass and rapidity of the colorless final state.
The sum runs over all flavor combinations $(i,j)$ contributing at Born level, $i j \to h$,
and $\sigma_0$ is the corresponding Born partonic cross section.%
\footnote{For ease of notation, we suppress the possibility of $\sigma_0$ depending on the flavors $i,j$.}
The hard function $H_{ij}$ encodes virtual corrections to the Born process $ij\to h$, \emph{i.e.}\ it is given as the corresponding renormalized form factor.
The beam functions $B_i(x,\Tau)$ encode the probability to extract a parton of type $i$ with momentum fraction $x$ from the proton, together with the contribution from collinear radiation to the observable $\Tau$,
while the soft function $S(\Tau)$ encodes the effect of soft exchange between the protons.
Since $S(\Tau)$ only differs between quark- and gluon-induced processes, we suppress an explicit flavor label.
Finally, $\otimes$ denotes a convolution in $\Obs$, whose precise structure depends on the chosen observable $\Obs$, and often can be made multiplicative in a suitable conjugate space.
Note that in \eq{fact_thm} we suppress explicit renormalization scales that are present in all functions.

The factorisation of degrees of freedom at the Lagrangian level makes the ingredients for the various functions in \eq{fact_thm} evident.
The hard function, $n$-collinear and $\bn-$collinear beam functions and the soft function are each defined in terms of only hard, $n$-collinear, $\bn$-collinear  and soft degrees of freedom, respectively.
This implies that the expansion techniques developed in this article are perfectly suited to determine beam functions from a perturbative computation using a pure collinear expansion of cross sections.
Here, it is important that both real and loop momenta are expanded as collinear.
We will provide explicit examples by obtaining the NNLO beam functions for $\Obs=q_T$ and $\Obs=\Tau_N$ ($N$-jettiness) in \sec{beamfunctions}.
We also note that in a similar fashion, one can also obtain the soft function by considering a purely soft expansion.

We also stress that since SCET is an effective field theory, it can be systematically extended by including the power-suppressed Lagrangians $\cL^{(k>0)}$ in \eq{LSCET}. This is the EFT analog of expanding cross sections to subleading order in $\lambda$ about the soft and collinear limits.
However, at subleading powers, collinear and soft interactions do not simply factorize similar to \eq{fact_thm} anymore, and factorization theorems and the resummation of large logarithms become much more involved~\cite{Hill:2004if,Lee:2004ja,Benzke:2010js,Freedman:2014uta,Moult:2016fqy,Moult:2017jsg,Beneke:2017vpq,Beneke:2017ztn,Feige:2017zci,Moult:2017rpl,
Chang:2017atu,Moult:2017xpp,Alte:2018nbn,Beneke:2018gvs,Beneke:2018rbh,Moult:2018jjd,Ebert:2018lzn,Ebert:2018gsn,Bhattacharya:2018vph,Beneke:2019kgv,Moult:2019mog,Beneke:2019mua,Moult:2019uhz,Moult:2019vou,Liu:2019oav}. 
Since our expansion technique allows us to perform collinear expansions of partonic cross sections to arbitrary order in $\lambda$, we hope that it will also provide insights into the structure of factorisation theorems beyond the leading power, and that it can be used to determine universal quantities like generalizations of soft and beam functions at subleading power.

%===============================================================================
\subsection{Soft-collinear overlap and zero-bin subtractions}
\label{sec:zero_bin}
%===============================================================================

In order to obtain a full description of a cross section in its infrared limit, we need to combine all collinear and soft regions. Schematically, we expand
\begin{align} \label{eq:zero_bin_1}
 \lim_{\rm IR} \frac{\sigma}{\df Q^2 \df Y \df \Obs} &
 = \frac{\sigma^{(n)}}{\df Q^2 \df Y \df \Obs} + \frac{\sigma^{(\bn)}}{\df Q^2 \df Y \df \Obs} + \frac{\sigma^{(s)}}{\df Q^2 \df Y \df \Obs} + \cdots
\,,\end{align}
where the $\sigma^{(n,\bn,s)}$ correspond to the expansion of the cross sections where all emissions are treated  as $n$-collinear, $\bn$-collinear and soft, respectively.
The ellipses denote mixings of these cases, as well as power-suppressed corrections.
Note that here in the following, we do not consider the hard region. While it is required to obtain an infrared-finite cross section, it corresponds to physics at the hard scale $\mu^2 \sim Q^2$, and does not affect the soft-collinear overlap discussed in the following.

In practice, \eq{zero_bin_1} is often too naive, as there is a nontrivial overlap between the collinear and soft regions. This arises because the soft limit of a squared matrix element is equal to the soft limit of the \emph{collinear limit} of the same matrix element.
As discussed and illustrated in more detail in \sec{soft_limit}, this can be understood since the soft limit can be equivalently obtained by either directly rescaling
\begin{align} \label{eq:soft_1}
 k^\mu = (k^+,k^-,k_\perp)
 \quad\stackrel{\mathrm{soft}}{\xrightarrow{\hspace*{1.5cm}}} \quad
 (\lambda^2, \lambda^2, \lambda^2)
\,,\end{align}
or by first rescaling into the collinear limit with a subsequent soft rescaling,
\begin{align} \label{eq:soft_2}
 k^\mu = (k^+,k^-,k_\perp)
 \quad\stackrel{n-\mathrm{collinear}}{\xrightarrow{\hspace*{1.5cm}}} \quad
 (1, \lambda^2, \lambda)
 \quad\stackrel{\mathrm{soft}}{\xrightarrow{\hspace*{1.5cm}}} \quad
 (\lambda^2, \lambda^2, \lambda^2)
\,.\end{align}
Since the second rescaling only lowers the scaling of each component, no information is lost, and \eqs{soft_1}{soft_2} produce the same expansion of a matrix element.

Consequently, when one integrates over a collinearly-rescaled momentum, the integral will always contain contributions from the soft region. Schematically, if we write
\begin{align}
 \int\df^d p \, f(p^+, p^-, p_\perp)
 \quad\stackrel{n-\mathrm{coll}}{\xrightarrow{\hspace*{1.cm}}}\quad
 \lambda^d \!\int\!\df p^+ \df p^- \df^{d-2} \vec p_\perp \, f^{(n)}(p^+ \sim \lambda^2, p^- \sim 1, p_\perp \sim \lambda)
\end{align}
for the collinear expansion $f^{(n)}$ of an arbitrary integrand $f$, then clearly the integration range extends into a region where the assumed collinear scaling is not justified.
In particular, the $p^-$ integral extends to $p^- \to 0$, which corresponds to a soft region.
This contribution to the soft region can be identified and extracted by further expanding $f^{(n)}$ as indicated in \eq{soft_2}.

In conclusion, the collinear limit of the cross section has an overlap with the soft limit, which can be extracted by an additional reexpansion in the soft limit, which has been demonstrated explicitly for a mixed real-virtual integral in \sec{soft_limit}.
We thus need to modify \eq{zero_bin_1} as
\begin{align} \label{eq:zero_bin_2}
  \lim_{\rm IR} \frac{\sigma}{\df Q^2 \df Y \df \Obs} &
 = \left[ \frac{\sigma^{(n)}}{\df Q^2 \df Y \df \Obs} - \frac{\sigma^{(n \to s)}}{\df Q^2 \df Y \df \Obs}\right]
 + \left[ \frac{\sigma^{(\bn)}}{\df Q^2 \df Y \df \Obs} - \frac{\sigma^{(\bn \to s)}}{\df Q^2 \df Y \df \Obs}\right]
 \nn\\&\quad
 + \frac{\sigma^{(s)}}{\df Q^2 \df Y \df \Obs} + \cdots
\,,\end{align}
where the soft limit of the collinear cross sections are denoted by $\sigma^{(n \to s)}$ and $\sigma^{(\bn \to s)}$, respectively. The terms in brackets hence correspond to the true $n$- and $\bn$-collinear limits of the cross section.
Note that in general, $\sigma^{(n \to s)} \ne \sigma^{(s)}$, because the observable $\Tau$ itself has to be expanded in the collinear and soft limits.

Let us connect these observations to the corresponding treatment in SCET.
As a modal EFT, SCET is built to separately describe soft and collinear modes, and hence as a matter of principle collinear momenta are not allowed to overlap with the soft sector.
In practice, it is not feasible to introduce a cutoff between soft and collinear modes.
Instead, one follows the same strategy outlined above: after calculating a collinear integral, one subtracts its soft limit to obtain the pure collinear result.
This procedure is referred to as zero-bin subtraction \cite{Manohar:2006nz}, and is crucial to a well-defined separation of modes in the EFT.
In practice, the zero-bin subtractions are often absent in dimensional regularization or equal to the soft function itself, and thus can be easily taken into account.

%% file: Chapters/BeamFunctions.tex
%%%%%%%%%%%%%%%%%%%%%%%%%%%%%%%%%%%%%%%%%%%%%%%%%%%%%%%%%%%%%%%%%%%%%%%%%%%%%%%%
\section{Beam functions from the collinear limit}
\label{sec:beamfunctions}
%%%%%%%%%%%%%%%%%%%%%%%%%%%%%%%%%%%%%%%%%%%%%%%%%%%%%%%%%%%%%%%%%%%%%%%%%%%%%%%%

In this section we show how the collinear expansions can be used to compute beam functions. We briefly introduce the notion of beam functions in \sec{beamdef}, and then show in \sec{beam_funcs_strategy} how they are related to the collinear expansion of cross sections developed before. Our method is briefly contrasted with other methods of calculating beam functions in \sec{beam_func_methods}. We show explicitly how to obtain the $N$-jettiness and the $q_T$ beam functions at NNLO using this method in \sec{Tau_beam_func} and \sec{qT_beam_func}, respectively.

%===============================================================================
\subsection{Beam functions}
\label{sec:beamdef}
%===============================================================================

Beam functions are defined as gauge-invariant hadronic matrix element that measure the large lightcone momentum entering the hard interaction, as well as the contribution to the observable $\Tau$ from collinear radiation.
For example, the quark beam function $B_q$ is defined in SCET as~\cite{Stewart:2009yx}
\begin{align} \label{eq:beam_def}
 B_q(x = p^-/P^-, \Tau) = \big< p_n(P) \big| \bar\chi_n(0) \frac{\slashed \bn}{2} \bigl[ \delta(p^- - \bn \cdot \cP) \, \delta(\Tau - \hat \Tau) \, \chi_n(0) \bigr] \big| p_n(P) \bigr>
\,.\end{align}
Here, the $\chi_n = W_n^\dagger q$ are collinear quark fields defined in SCET as quark fields dressed with collinear Wilson lines $W_n$, $p_n(P)$ is a proton moving along the $n$-direction with momentum $P$, and $\bn\cdot\cP$ is the SCET momentum operator that determines the lightcone momentum of all fields to its right.
By boost invariance, the beam function only depends on the momentum fraction $x = p^-/P^-$.
Similarly, $\hat\Tau$ is the appropriate measurement operator determining the observable $\Tau$ in terms of all momenta of the fields to its right.

Beam functions are a natural generalization of PDFs, which in SCET are defined as~\cite{Bauer:2002nz}
\begin{align} \label{eq:pdf_def}
 f_q(x = p^-/P^-) = \bigl<p_n(P) \big| \bar\chi_n(0) \frac{\slashed n}{2} \bigl[ \delta(p^- - \bn \cdot \cP) \, \chi_n(0) \bigr] \big| p_n(P)\bigr>
\,.\end{align}
Comparing \eqs{beam_def}{pdf_def}, it is evident that the beam function extends the PDF by measuring an additional observable $\Tau$ on top of the longitudinal momentum fraction carried by the struck parton.

Both beam functions and PDFs are in general intrinsically nonperturbative matrix elements.
For perturbative $\Tau \gg \lqcd$, one can perform an operator product expansion of the beam function onto the PDF~\cite{Stewart:2009yx},
\begin{align} \label{eq:matching_B}
 B_i(x,\Tau,\mu) = \sum_j \, \cI_{ij}(x,\Tau,\mu)\otimes_x  f_j^R(x,\mu) \times \bigl[1 + \cO(\lqcd/\Tau)\bigr]
\,.\end{align}
Here, the only nonperturbative input is given in terms of the PDFs, while the matching kernel $\cI_{ij}$ are perturbatively calculable.

For completeness, we remark that PDFs and beam functions can also be defined without invoking SCET by expressing the collinear quark fields $\chi_n$ in terms of standard quark fields and collinear Wilson lines $W_n$, which are defined as path-ordered exponentials of the gluon field projected onto the appropriate collinear direction. Beam functions are also often written as the Fourier transform of a position-space correlator, where the separation between the quark fields corresponds to the exchanged momentum and often avoids the need for the momentum operator $\cP$ in \eq{beam_def}. PDFs and TMDPDFs were originally defined in this way~\cite{Collins:1981uw,Collins:1984kg}, and the equivalence of both formulations was also discussed in the context of $\Tau_N$ beam functions in \refscite{Stewart:2009yx,Stewart:2010qs}.
Note that the study of parton distributions from lattice QCD requires the definition in position space, see e.g.~\refscite{Ji:2014hxa,Ji:2018hvs,Ebert:2018gzl,Ebert:2019okf,Ebert:2019tvc,Ji:2019sxk,Ji:2019ewn,Vladimirov:2020ofp,Ebert:2020gxr} for recent progress towards calculating TMDPDFs on lattice, and \refscite{Cichy:2018mum, Ji:2020ect} for a more general overview of parton physics from lattice QCD. For perturbative calculations, both formulations are equivalent.

%===============================================================================
\subsection{General strategy}
\label{sec:beam_funcs_strategy}
%===============================================================================

In \sec{facthm}, we discussed that the hard, beam and soft functions in the factorized cross section in \eq{fact_thm} are each defined only in terms of the hard, collinear and soft modes of \eq{modes}, respectively. Hence, in the limit where all loop and final-state momenta are treated as $n$-collinear, the hard function, the $\bn$-collinear beam function, and the soft function only contribute at their respective tree level, where they are normalized to unity and flavor diagonal. Thus, the strict $n$-collinear limit of \eq{fact_thm} is given by
\begin{equation} \label{eq:fact_thm_n}
\strictlim \frac{\df\sigma}{\df Q^2 \df Y \df \Tau}
 = \sigma_0 \sum_{i,j} B_i(x_1^B,\Tau) f_j(x_2^B)
\,,\end{equation}
where we remind the reader that the flavor sum runs over all flavors contributing at Born level, $i j \to h$, and $\sigma_0$ is the associated Born partonic cross section.

We remark that \eq{fact_thm_n} is to be understood at the bare level, as it for example does not encode scale independence. Indeed, as we will see, even after UV renormalization and IR subtraction one encounters leftover poles in $\eps$, which in the full factorized cross section in \eq{fact_thm} cancel with the other ingredients.

In the following, we assume that the Born process is diagonal in flavor, \emph{i.e.} only the $gg$ channel (as in Higgs production in gluon fusion) or the $q\bar q, \bar q q$ channels (as in Drell-Yan or $b\bar{b}$ initiated Higgs production) contribute, where $q$ is an arbitrary quark flavor. With this assumption,  we can fix $j = \bar i$ in \eq{fact_thm_n}, which allows us to easily read off the \emph{bare} beam function by comparing with the $n$-collinear limit of the cross section given in \eqs{sigma_hadr}{partcoef_special},
\begin{align} \label{eq:beam_master}
 B_i(x_1^B,\Tau) &
 = \sum_{j} \int_{x_1^B}^1 \frac{\df z_1}{z_1} f_j\Bigl(\frac{x_1^B}{z_1}\Bigr)
   \times \int_0^1 \df x \int_0^\infty \df \wa\df \wb \, \delta\left[z_1-(1-\wa)\right]
   \nn\\&\qquad\times
   \strictlim \left\{\delta\left[\Obs-\Obs(Q,Y,\wa,\wb,x)\right] \frac{\df\eta_{j \bar i}}{ \df Q^2  \df \wa \df \wb \df  x}\right\}
\,.\end{align}
By fixing the flavor of the $\bn$-collinear parton as $\bar i$, we extract the correct beam function for the flavor $i$ in a flavor-diagonal process.

\Eq{beam_master} has precisely the structure of \eq{matching_B}, and we can immediately read off the bare matching kernel as the collinear limit of the partonic coefficient function,
\bea \label{eq:I_bare}
 \cI_{ij}^{\rm bare}(z,\Tau)&=&
   \int_0^1 \df x \int_0^\infty \df \wa\df \wb \, \delta\left[z-(1-\wa)\right]  \nonumber\\
&\times& \strictlim \left\{\delta\left[\Obs-\Obs(Q,Y,\wa,\wb,x)\right] \frac{\df\eta_{j\bar i}}{ \df Q^2  \df \wa \df \wb \df  x}\right\}
\,.\eea
We stress that the partonic coefficient function here is limited to strictly collinear modes only.
In contrast, in the collinear expansion for cross sections discussed before, we also included non-collinear modes when computing loop integrals.
However, we showed that the collinear and non-collinear modes can easily be separated by looking at their respective generalized scaling behaviour.
Extracting the required parts is consequently easy. 
In the strictly collinear limit the general partonic coefficient function of \eq{eta_ij_1} becomes
\begin{align}
 &\strictlim \frac{\df\eta_{j\bar i} }{\df Q^2 \df w_1 \df w_2 \df x}
\\\nn&
 = \delta_{j\bar i} \delta(w_1)\delta(w_2)\delta(x)
 \,+\, \sum_{\ell=1}^\infty \left(\frac{\as}{\pi}\right)^{\ell} w_2^{-1-l\eps}
   \sum_{m=1}^\ell w_1^{-1-m\eps}
   \frac{\df\eta_{j\bar i}^{(\ell,m,n)}(w_1,0,x,Q^2)}{\df Q^2 \df w_1 \df w_2 \df x}
\,.\end{align}
The strict collinear limit for the partonic coefficient function for the observable $\Tau$ in \eqs{beam_master}{I_bare} is then obtained in analogy to \eq{partoniccoefexp}.

A special case of \eq{I_bare} is the bare matching kernel differential in $\wa$, $\wb$ and $x$ itself, from which one can project out all other beam functions we are interested in.
In fact, this double-differential beam function can be related to the fully unintegrated parton distribution first formulated in \refscite{Collins:2007ph,Rogers:2008jk} and within SCET in \refscite{Mantry:2009qz,Jain:2011iu}, where it is also known as double-differential beam function (dBF).
Importantly, in general the projection of $(\wa,\wb,x)$ onto the desired observable $\Tau$ only holds at the bare level, not after renormalization of the dBF~\cite{Jain:2011iu}.
The renormalization of the dBF is also significantly more complicated than that of the $\Tau_N$ and $q_T$ beam functions we are interested in, see \refscite{Gaunt:2014xxa,Gaunt:2020xlc} for explicit results at NNLO.

$\cI_{ij}^{\rm bare}$ still contains infared poles that cancel upon PDF renormalization in \eq{beam_master}.
Even after $\as$ renormalization, this still leaves divergences that cancel in the cross section when combining the $n$-collinear limit with the $\bn$-collinear and soft limits, but are remnant in the bare matching kernel.
In the EFT, these divergences are of ultraviolet origin and thus can be absorbed in the standard fashion through a counterterm.
Subtracting both IR and UV poles in this manner, we obtain the renormalized matching kernel as
\begin{equation} \label{eq:I_master}
 \cI_{ij}(x,\Tau,\mu)
 = \sum_{j'} \Gamma_{j j'}(z, \eps) \otimes_z Z_B^i (\Tau,\mu,\eps) \otimes_\Tau \hat Z_{\as}(\mu,\eps) \, \cI_{ij'}^{\rm bare}(x,\Tau,\eps)
\,,\end{equation}
where $\otimes_\Tau$ denotes the appropriate convolution in $\Tau$.
According to \eqs{I_bare}{I_master}, we can obtain the beam function matching kernel as follows:
\begin{enumerate}
 \item Obtain the bare kernel $\cI_{ij}^{\rm bare}$ from the strict collinear limit of the partonic cross section.
 \item Apply $\alpha_s$ renormalization through $\hat Z_{\as}$, which renormalizes the bare coupling constant $\as^b$ in the $\MSbar$ scheme.
 \item Subtract the EFT UV divergences with the beam-function counterterm $Z_B^i$.
       This renormalization does not change the parton flavor $i$, and only differs between quark and gluons, but is independent of the quark flavor.
       In general, this counterterm enters through a convolution in $\Tau$, which can be trivialized by going to suitable conjugate space.
 \item Subtract IR divergences by convolving with the PDF counterterm $\Gamma_{jj'}$, which as usual mixes parton flavors.
\end{enumerate}
Since the $\Gamma_{jj'}$ and $Z_B^i$ commute, one can freely rearrange their order in \eq{I_master}.
Since the beam function counter term $Z_B^i$ gives rise to the renormalization group equation of the beam function, in practice one can either predict $Z_B^i$ from the RGEs and check that this cancels all poles in $\eps$, or determine $Z_B^i$ by absorbing all poles remaining after QCD UV and IR subtraction and verify that it reproduces the RGE dictated by the EFT. For the $\Tau_N$ and $q_T$ beam functions, this is discussed in more detail in our companion papers \cite{Ebert:2020yqt,Ebert:2020unb}.

In \eq{I_bare}, we assumed that the partonic coefficient function is taken in the strict $n$-collinear limit.
As discussed in \sec{zero_bin}, for certain observables there can be overlap with the soft limit, which in the factorized cross section in \eq{fact_thm} is already accounted for by the soft function.
In such instances, one has to subtract off the soft-collinear overlap,
\begin{align} \label{eq:I_bare_0bin}
 \cI_{ij}^{\rm bare}(z,\Tau) &
 = \int_0^1 \df x \int_0^\infty \df \wa\df \wb \, \delta\left[z-(1-\wa)\right]
 \nn\\&\quad\times
 \left[\strictlim\frac{\df \eta_{j\bar{i}}}{\df Q^2 \df Y \df \Tau} - \slim \strictlim\frac{\df \eta_{j\bar{i}}}{\df Q^2 \df Y \df \Tau} \right]
\,.\end{align}
The second term in the above equation denotes that the collinear limit is further re-expanded in the soft limit.

%===============================================================================
\subsection{Comparison to alternative methods}
\label{sec:beam_func_methods}
%===============================================================================

Before illustrating our method for the $\Tau_N$ and $q_T$ beam functions in \secs{Tau_beam_func}{qT_beam_func}, we briefly contrast our approach to methods previously used in the literature.
Here, we focus on how to calculate the bare beam function, since the renormalization and subtraction of UV and IR divergences always proceeds in the same fashion.

Most calculations of beam functions explicitly calculate matching coefficients from matrix element of the beam function operator, see e.g.~\refscite{Stewart:2010qs,Gaunt:2014xga,Gaunt:2014cfa,Gaunt:2014xxa,Gehrmann:2012ze,Gehrmann:2014yya,Gangal:2016kuo,Echevarria:2016scs,Gutierrez-Reyes:2019rug,Luo:2019hmp,Luo:2019bmw,Luo:2019szz,Gaunt:2020xlc}. Let us explain some features of this approach for the concrete example of a quark beam function as shown in \eq{beam_def}, whose bare matching kernel $\cI_{qj}$ is obtained by evaluating the matrix element in \eq{beam_def} with an external on-shell parton of flavor $j$. In \refcite{Stewart:2010qs}, the analytic structure of these matrix elements was discussed in detail for the $\Tau_N$ beam function, and it was shown that one can calculate it by taking the discontinuity of matrix elements of the time-ordered operator.

Firstly, this implies that the beam function can be calculated using SCET Feynman rules.
Since a single collinear sector in SCET is equal to a boosted copy of QCD, one can equivalently employ QCD Feynman rules. In this case, eikonal vertices arise from the Feynman rules of the Wilson lines $W_n$ that are part of the collinear quark fields $\chi_n = W_n^\dagger q$. These can be avoided in lightcone gauge, where $\bn \cdot A = 0$ such that $W_n = 1$, but similar terms arise from the gluon propagator in ligthcone gauge. Since the beam function is defined as a gauge-invariant matrix element, both approaches yield equal results.

The discontinuity can be obtained by using the Cutkosky rules~\cite{Cutkosky:1960sp} (see also \refcite{Ellis:1996nn}), which corresponds to taking particles exchanged between the quark fields in \eq{beam_def} on-shell. This is analogous to our approach, where we explicitly consider on-shell radiation into the final state.
Alternatively, one can not apply an on-shell constraint and integrate over all particles, and explicitly take the discontinuity afterwards. Both approaches are discussed in more detail in \refcite{Gaunt:2014cfa}, where they are referred to as on-shell and dispersive method, respectively.

An alternative method that does not directly rely on the definition of the beam function in SCET was pointed out in \refcite{Ritzmann:2014mka}, where it was shown that one can equivalently calculate the beam function from phase-space integrals over QCD splitting functions. This approach was used in \refscite{Baranowski:2020xlp,Melnikov:2018jxb,Melnikov:2019pdm,Behring:2019quf}, where the required splitting function at N$^3$LO was obtained following the method of \refcite{Catani:1999ss}. This approach requires to use a physical gauge where gluons are explicitly transverse, for example the lightcone gauge $\bn \cdot A = 0$.

Similar to \refcite{Ritzmann:2014mka}, our method does not rely on directly calculating SCET matrix elements.
However, our approach is manifestly gauge invariant as it is based on a physical cross section, similar to the direct calculations.
The connection of our approach to these previous methods can be understood as follows: 
Prior to integrating over real radiation, the collinear expansion reproduces precisely the collinear limit of QCD, which in the SCET approach is immediately encoded in the structure of the SCET matrix element, whereas in the approach of \refcite{Ritzmann:2014mka} it is obtained from the QCD splitting function.
In practice, one advantage of our method is that it can be easily integrated with standard methods of generating Feynman diagrams.
One can then use standard methods to evaluate the integrals using IBPs~\cite{Chetyrkin:1981qh,Tkachov:1981wb} and the method of differential equations~\cite{Kotikov:1990kg,Kotikov:1991hm,Kotikov:1991pm,Henn:2013pwa,Gehrmann:1999as} in the reverse unitarity framework~\cite{Anastasiou2003,Anastasiou:2002qz,Anastasiou:2003yy,Anastasiou2005,Anastasiou2004a} over the real radiation phase space, keeping only the total momentum $k$ fixed. This intermediate result, $\df\eta_{ij} / (\df Q^2 \df\wa \df\wb \df x)$, is the bare fully differential beam function, from which one can then project out the desired beam functions.

%===============================================================================
\subsection[\texorpdfstring{$\Tau_N$}{TauN} beam functions]
           {\boldmath $\Tau_N$ beam functions}
\label{sec:Tau_beam_func}
%===============================================================================

%===============================================================================
\subsubsection{Factorization}
\label{sec:tau0_factorization}
%===============================================================================

$N$-jettiness is an inclusive event shape that yields an $N$-jet resolution variable.
It was first introduced in \refcite{Stewart:2010tn}, and its factorization was derived using SCET in \refscite{Stewart:2009yx, Stewart:2010tn, Jouttenus:2011wh}.
Since the same beam function appears for all $\Tau_N$, we focus only on the simplest case $\Tau_0$, also known as beam thrust, that is relevant to color-singlet processes.
Beam thrust is defined as \cite{Stewart:2010tn, Jouttenus:2011wh}
\begin{align} \label{eq:Tau0}
\Tau_0 = \sum_i {\rm min}\biggl\{ \frac{2 q_1 \cdot (-k_i)}{Q_a},\, \frac{q_2 \cdot (-k_i)}{Q_b} \biggr\}
\,.\end{align}
Here, $q_{1,2}$ are the Born-projected momenta of the incoming partons, given by
\begin{align}
 q_1^\mu = x_1^B \sqrt{S} \frac{n^\mu}{2} = Q e^Y \frac{n^\mu}{2}
\,,\qquad
 q_2^\mu = x_2^B \sqrt{S} \frac{\bn^\mu}{2} = Q e^{-Y} \frac{\bn^\mu}{2}
\,,\end{align}
where as before $Q$ and $Y$ are the invariant mass and rapidity of the color-singlet final state $h$, respectively.
The sum in \eq{Tau0} runs over all final-state particles excluding $h$, and as usual all final-state momenta are taking as incoming.
The $Q_{a,b}$ are measures that determine different definitions of 0-jettiness.
The original definitions are~\cite{Stewart:2009yx, Berger:2010xi}
\begin{alignat}{4} \label{eq:Tau0_2}
 &\text{leptonic:}\quad & Q_a &= Q_b = Q \,,\qquad
  & \Tau_0^{\rm lep} &= -\sum_i \min \Bigl\{ e^Y n \cdot k_i \,,\, e^{-Y} \bn \cdot k_i \Bigr\}
\nn\\
 &\text{hadronic:}\qquad & Q_{a,b} &= Q \, e^{\pm Y} \,,
 & \Tau_0^{\rm cm} &= -\sum_i \min \Bigl\{ n \cdot k_i \,, \bn \cdot k_i \Bigr\}
\,.\end{alignat}
The precise choice does not affect the calculation of the beam function, but it becomes important for the calculation of power corrections~\cite{Ebert:2018lzn}.
We note in passing that at subleading power, the leptonic definition is clearly preferred as it gives rise to smaller power corrections that the hadronic definition \cite{Moult:2016fqy,Ebert:2018lzn}.

At small $\Tau_0 \ll Q$, the cross section can be factorized as~\cite{Stewart:2009yx}
\begin{align} \label{eq:Tau0_fact}
 \frac{\df\sigma}{\df Q^2 \df Y \df \Tau_0} &
= \sigma_0 \sum_{i,j} H_{ab}(Q^2, \mu) \int \! \df t_a \, \df t_b \,
   B_a(t_a, x_1^B, \mu) \, B_b(t_b, x_2^B, \mu) \,
   S\Bigl(\Tau_0 - \frac{t_a}{Q_a} - \frac{t_b}{Q_b}, \mu\Bigr)
\nn \\ &\qquad
\times \Bigl[1 + \cO\Bigl(\frac{\Tau_0}{Q}\Bigr)\Bigr]
\,.\end{align}
As indicated, this factorization holds up to power corrections suppressed by $\Tau_0/Q$ that were studied in~\refscite{Moult:2016fqy,Moult:2017jsg,Boughezal:2018mvf,Ebert:2018lzn,Boughezal:2019ggi} and the relevant SCET operators have been derived in \refscite{Feige:2017zci,Moult:2017rpl,Chang:2017atu}.
In the case of fiducial cuts applied to the decay products of $h$, these corrections can be enhanced as $\cO(\sqrt{\Tau_0/Q})$ \cite{Ebert:2019zkb}.
Furthermore, starting at N$^4$LO it also receives contributions from perturbative Glauber-gluon exchanges that are not captured by \eq{Tau0_fact}~\cite{Gaunt:2014ska, Zeng:2015iba}.

The beam function $B_i(t,x,\mu)$, sometimes also referred to as the virtuality-dependent beam function, appears in the factorization of all $\Tau_N$~\cite{Stewart:2010tn}, deep-inelastic scattering \cite{Kang:2013nha}, and in the factorization of color-singlet processes in the generalized threshold limit \cite{Lustermans:2019cau}.
It is known at NNLO~\cite{Stewart:2010qs,Berger:2010xi,Gaunt:2014xga,Gaunt:2014cfa}, and we compute it at N$^3$LO for all partonic channels in our companion paper~\cite{Ebert:2020unb}. Previous progress towards the calculation of the quark beam function at N$^3$LO was made in \refscite{Melnikov:2018jxb,Melnikov:2019pdm,Behring:2019quf}.

In \eq{Tau0_fact}, the beam functions are defined to measure the $Q_{a,b}$-independent combinations $t_a = -q_1^ -k^+$ and $t_b = - q_2^+ k^-$, while the measurement-dependent normalization factors $Q_{a,b}$ only arise in the convolution in \eq{Tau0_fact}.
This definition naturally arises because $\Tau_0$ simplifies in the $n$-collinear limit to
\begin{align} \label{eq:Tau0_n}
 \nlim \Tau_0 = \sum_i \frac{2 q_1 \cdot (-k_i)}{Q_a} = \frac{q_1^- (-k^+)}{Q_a}
\,,\end{align}
and similarly in the $\bn$-collinear limit.

The soft function in \eq{Tau0_fact} only differs between quark annihilation and gluon fusion, but is independent of quark flavors, and we suppress the explicit color index in \eq{Tau0_fact}.
$S(\Tau,\mu)$ is a hemisphere soft function for two incoming lightlike Wilson lines.
Through NNLO, it is equal to the hemisphere soft function for $e^+ e^- \to \rm{dijets}$ \cite{Stewart:2009yx,Kang:2015moa}, which itself is known at NNLO  \cite{Stewart:2009yx,Stewart:2010qs,Schwartz:2007ib,Fleming:2007xt,Kelley:2011ng,Monni:2011gb,Hornig:2011iu}.

%===============================================================================
\subsubsection[Calculation of \texorpdfstring{$\Tau_N$}{TauN}-dependent beam functions]
              {\boldmath Calculation of $\Tau_N$-dependent beam functions}
\label{sec:tau0_beam_funcs}
%===============================================================================

Since the collinear limit of $\Tau_0$ given in \eq{Tau0_n} only depends on the total momentum $k^\mu$ of all real emissions, the $\Tau_N$ beam function can be calculated using the method outlined in \sec{beam_funcs_strategy}. In contrast, the soft limit of \eq{Tau0} requires knowledge of all individual momenta $\{k_i\}$, and thus can not be calculated in this fashion.

Using \eqs{I_bare}{Tau0_n}, we can calculate the bare beam function kernel as
\begin{align} \label{eq:cI_tau_bare}
 \cI_{ij}^{\rm bare}(z,t,\eps) &
 = \int_0^1 \df x \int_0^\infty \df w_1  \df w_2
   \, \delta[z-(1-\wa)] \, \delta\bigl(t - Q^2 w_2 \bigr)
   \strictlim\frac{\df\eta_{j\bar{i}} }{\df Q^2 \df w_1 \df w_2 \df x}
\,.\end{align}
The zero-bin for the $\Tau_N$ beam function is known to be scaleless and thus vanishes in pure dimensional regularization, and hence need not be included explicitly \cite{Stewart:2010tn}.
Note that $w_2 > 0$ implies $t > 0$, which we keep implicit. 

The bare kernel contains UV divergences from the limit $\wa=1-z\to0$ and $\wb=t/Q^2\to0$, which are both regulated using dimensional regularization. The divergences from small $t$ can be made manifest through the standard identity
\begin{align} \label{eq:distr_tau}
 \frac{1}{\mu^2} \left(\frac{\mu^2}{t}\right)^{1 + a \eps} = - \frac{\delta(t)}{a \eps} 
 + \left[\frac{1}{t}\right]_+ + a \eps \left[\frac{\ln(t/\mu^2)}{t}\right]_+  + \cO[(a \eps)^2]
\,,\end{align}
where $\left[\ln^n x / x\right]_+$ is the standard plus distributions.
Following \sec{beam_funcs_strategy}, we obtain the renormalized matching as
\begin{equation} \label{eq:Iij_ren}
 \cI_{ij}(t,z,\mu)
 = \sum_{k} \int\df t' \, Z_B^i(t-t',\eps,\mu) \, \int_z^1\!\frac{\df z'}{z'}
   \, \Gamma_{jk}\Bigl(\frac{z}{z'},\eps\Bigr) \,
   \hat Z_{\as}(\mu,\eps) \, \cI_{i k}(t',z',\eps)
\,,\end{equation}
where the structure of convolution in $t$~\cite{Stewart:2010qs,Berger:2010xi} is made explicit.
In practice, it is more useful to perform the renormalization in Fourier or Laplace space, where the convolution in $t$ turns into a simple product. In particular, the structure of $Z_B^i$ can be easily predicted from the beam function RGE in Fourier space. For details on this, we refer to \refcite{Ebert:2020unb}.

We have implemented the described procedure at one loop through $\cO(\eps^4)$ and at two loops through $\cO(\eps^2)$, as required for the calculation of the three-loop beam function.
We use the collinear limit of the cross sections for Higgs and Drell-Yan production to extract the gluon and quark beam functions, respectively.
As intermediate checks, we verified that the UV and IR counterterms correctly cancel all appearing divergences.
The final renormalized results agrees with the NNLO results reported in \refscite{Gaunt:2014xga,Gaunt:2014cfa},
and the higher-order terms in $\eps$ agree with \refcite{Baranowski:2020xlp}.
Our bare results are provided as ancillary files.

%===============================================================================
\subsection[\texorpdfstring{$q_T$}{qT} beam functions]
           {\boldmath $q_T$ beam functions}
\label{sec:qT_beam_func}
%===============================================================================

%===============================================================================
\subsubsection{Factorization}
\label{sec:qT_factorization}
%===============================================================================

The factorization of the transverse-momentum ($\qt$) distribution of a colorless probe $h$ in the limit $q_T \ll Q$ was first derived by Collins, Soper, and Sterman (CSS) in \refscite{Collins:1981uk,Collins:1981va,Collins:1984kg} and elaborated on in \refscite{Catani:2000vq, deFlorian:2001zd, Catani:2010pd, Collins:1350496}. The factorization was also discussed using SCET in \refscite{Becher:2010tm, GarciaEchevarria:2011rb, Chiu:2012ir, Li:2016axz}.
The factorized cross section is commonly formulated in Fourier (impact parameter) space, with $\bt$ being Fourier-conjugate to $\qt$, as this significantly simplifies the resummation of large logarithms~\cite{Ebert:2016gcn}. We write the factorized $\qt$ spectrum as
\begin{align} \label{eq:qt_fact}
 \frac{\df \sigma}{\df Q^2 \df Y \df^2 \qt} &
 = \sigma_0 \sum_{i,j} H_{ij}(Q^2,\mu) \int\!\df^2\bt \, e^{\img\,\qt \cdot \bt} \,
   \,\tilde B_i\Bigl(x_1^B, b_T, \mu, \frac{\nu}{\omega_a}\Bigr)
   \,\tilde B_j\Bigl(x_2^B, b_T, \mu, \frac{\nu}{\omega_b}\Bigr)
   \, \tilde S(b_T, \mu, \nu)
\nn\\&\quad
 \times \left[1 + \cO\left(q_T^2/Q^2\right) \right]
\,.\end{align}
It receives power corrections suppressed by $q_T^2/Q^2$, which were studied at fixed order in perturbation theory in \refcite{Ebert:2018gsn}. The study of their all-order structure has been initiated using the SCET operator formalism in \refscite{Kolodrubetz:2016uim,Feige:2017zci,Moult:2017rpl,Chang:2017atu,Chang:NLP},
and their nonperturbative structure has been explored in \refscite{Balitsky:2017gis,Balitsky:2017flc}.
These corrections are enhanced as $\cO(q_T/Q)$ when applying fiducial cuts to $h$ \cite{Ebert:2019zkb},
but for Drell-Yan and Higgs production can be uniquely included in the factorization theorem~\cite{Ebert:fiducial}, and are also linear when one includes radiation from massive final states \cite{Buonocore:2019puv}.

TMD factorization is complicated by the fact that the bare beam and soft functions not only contain IR and UV divergences,
but also so-called rapidity divergences. These must be regularized using a dedicated rapidity regulator,
and after removing the regulator this gives rise to the rapidity renormalization scale $\nu$.
Several such regulators are known in the literature \cite{Collins:1981uk,Collins:1350496,Becher:2010tm,Becher:2011dz,GarciaEchevarria:2011rb,Chiu:2011qc,Chiu:2012ir,Li:2016axz,Rothstein:2016bsq,Ebert:2018gsn}, leading to several equivalent schemes for defining TMD beam and soft functions.
It is also common to combine beam and soft functions into a $\nu$-independent TMDPDF as
\begin{equation}
 \tilde f_i(x,b_T,\mu,\zeta_i) = \tilde B_i(x, b_T, \mu, \nu/\sqrt\zeta_i) \sqrt{\tilde S(b_T,\mu,\nu)}
\,,\end{equation}
where $\zeta_i \propto \omega_i^2$ is known as the Collins-Soper scale \cite{Collins:1981va,Collins:1981uk}.

The TMD beam and soft functions appearing in \eq{qt_fact} are known at NNLO in various regulators \cite{Catani:2011kr,Catani:2012qa,Gehrmann:2012ze,Gehrmann:2014yya,Echevarria:2016scs,Gutierrez-Reyes:2019rug,Luo:2019hmp,Luo:2019bmw,Echevarria:2015byo,Luebbert:2016itl}.
The quark beam function and the soft function are also known at N$^3$LO \cite{Li:2016ctv,Luo:2019szz} using the exponential regulator of \refcite{Li:2016axz}.

An important remark is in order concerning differences between quark- and gluon-induced processes.
In the quark case, \eq{qt_fact} exactly applies, while the gluon beam functions can also depend on the gluon helicity due the vectorial nature of $\bt$.
As first pointed out in \refcite{Catani:2010pd}, the gluon beam function can be decomposed into a polarization-independent piece $B_1$ and a polarization-dependent piece $B_2$ as
\begin{align} \label{eq:Bg_decomposition}
 \tilde B^{\rho\lambda}_g(x,b_T,\mu,\nu)
 = \frac{g_\perp^{\rho\lambda}}{2} \tilde B_1(x,b_T,\mu,\nu)
 + \biggl(\frac{g_\perp^{\rho\lambda}}{2} - \frac{b_\perp^\rho b_\perp^\lambda}{b_\perp^2} \biggr) \tilde B_2(x,b_T,\mu,\nu)
\,.\end{align}
Here, $b_\perp^\mu$ is a Minkowski four vector with $b_\perp^2 = -b_T^2$, and $g_\perp^{\rho\lambda}$ is the transverse component of the metric tensor.
In this case, the hard function in \eq{qt_fact} also depends on the helicities of the colliding gluons.

We will only focus on the production of scalar particles such as a Higgs boson, where the hard function has the trivial helicity structure
\begin{align}
 H_{gg}^{\rho\lambda\rho'\lambda'}(Q,\mu) = H_{gg}(Q,\mu) g_\perp^{\rho\rho'} g_\perp^{\lambda\lambda'}
\,.\end{align}
Thus, the only combination that enters the factorized cross section in this case is
\begin{align} \label{eq:HBB_gg}
 H_{gg\,\rho\lambda\rho'\lambda'} \tilde B^{\rho\lambda}_g \tilde B^{\rho'\lambda'}_g
 = H_{gg} \tilde B^{\rho\lambda}_g \tilde B_{g\,\rho\lambda}
 = \frac12 H_{gg} \bigl[ \tilde B_1 \tilde B_1 + \tilde B_2 \tilde B_2 \bigr]
\,,\end{align}
where we suppress the arguments of all functions for brevity.
Since $\tilde B_2$ describes a spin flip of the incoming gluon, it vanishes at tree level, and thus the $\tilde B_2 \tilde B_2$ term first contributes at $\cO(\as^2)$.
Thus, for a scalar process, the $\tilde B_2 \tilde B_2$ term does not show up in the strict $n$-collinear limit, which hence can be used to calculate $\tilde B_1$ in the same fashion as for the quark case.
Nevertheless, $\tilde B_2$ could be calculated with the same technique for a different process that induces a cross term $\tilde B_1 \tilde B_2$, for example the production of a pseudoscalar probe $h$.
We also note that since $\tilde B_2$ is already known at NNLO \cite{Gutierrez-Reyes:2019rug,Luo:2019bmw}, the $\tilde B_2 \tilde B_2$ term in \eq{HBB_gg} is already known at N$^3$LO.

%===============================================================================
\subsubsection[Calculation of \texorpdfstring{$q_T$}{qT}-dependent beam functions]
              {\boldmath Calculation of $q_T$-dependent beam functions}
\label{sec:qT_beam_funcs}
%===============================================================================

In our setup for the differential hadronic cross section, \eq{sigma_differential}, we measured the transverse momentum of $h$ indirectly through
\begin{align} \label{eq:qT_measure}
 x = 1 - \frac{k_T^2}{k^+ k^-} = 1 - \frac{k_T^2}{s \wa \wb}
\,,\end{align}
as by momentum conservation $k_T = q_T$.
Both are defined as the magnitude of a $d-2$-dimensional vector, with the associated solid angle already integrated over in the phase space measure.
The $q_T$ measurement can also be defined in different schemes to account for extending the transverse vector into $d-2$ dimensions, but the scheme dependence must cancel in the renormalized beam functions.
For a more detailed discussion, see e.g.~\refcite{Luebbert:2016itl}.

Using \eqs{I_bare}{qT_measure} together with the leading-power relation $Q^2 = z s$, we obtain the matching kernel of the beam function as
\begin{align} \label{eq:cI_qT_naive}
 \cI_{ij}^{\rm naive}(z,q_T,\eps) &
  = \int_0^1 \df x \, \int_0^\infty \df w_1 \, \df w_2 \, \delta[z-(1-\wa)]
   \delta\biggl(q_T^2 - \frac{1-x}{z} w_1 w_2 Q^2 \biggr)
   \nn\\&\qquad\times
   \strictlim\frac{\df\eta_{j\bar{i}} }{\df Q^2 \df w_1 \df w_2 \df x}
\,.\end{align}
Here, the superscript $^{\rm naive}$ indicates that this is not yet the final result for the bare matching kernel, as it requires further manipulation.
First, we note that \eq{cI_qT_naive} contains divergences as $x\to1$ or $z\to1$ that are not regulated by dimensional regularization, and are a manifestation of the aforementioned rapidity divergences.
In our setup, we must regulate these with a regulator that acts only on the total radiation momentum $k^\mu$, but not on individual emissions. The only such regulator known in the literature is the exponential regulator of \refcite{Li:2016axz}, where one inserts a factor $\exp[2 \tau e^{-\gamma_E} k_i^0]$ into the phase of each real emission $k_i$. Inserting this regulator into \eq{cI_qT_naive} and solving the $\delta$ functions, we obtain
\begin{align} \label{eq:cI_qT_naive_2}
 \cI_{ij}^{\rm naive}(z,q_T,\eps,\tau/\omega) &
 = \lim_{\substack{\tau\to0\\\eps\to0}} \int_0^1 \df x \, \frac{1}{(1-x) (1-z)}
   \exp\biggl[-\tau e^{-\gamma_E} \frac{q_T^2}{\omega} \frac{z}{(1-z)(1-x)} \biggr]
   \nn\\&\hspace{1.2cm}\times
  \strictlim\frac{\df\eta_{j \bar{i}} }{\df Q^2 \df w_1 \df w_2 \df x}
   \bigg|_{ w_2 = \frac{q_T^2}{Q^2} \frac{z}{(1-x)(1-z)},\,w_1=1-z}
\,,\end{align}
where we defined the so-called label momentum of the beam function as $\omega = Q e^Y$.
In \eq{cI_qT_naive_2}, all divergences as $x\to1$ and $z\to1$ are manifestly regulated by the exponential, and any leftover divergences are regulated by dimensional regularization.
As indicated, the limit $\tau\to0$ should be taken before the limit $\eps\to0$.

To proceed, we Fourier transform to the conjugate $\bt$ space, which trades convolutions in $\qt$ for simple products in Fourier space. In $d-2$ dimensions, the Fourier transform reads
\begin{align} \label{eq:cI_qT_bare_3}
 \tilde \cI_{ij}^{\rm naive}(z,b_T,\eps,\tau/\omega)
 = \int\df^{d-2}\qt \, e^{-\img \bt \cdot \qt} \cI_{ij}^{\rm naive}(z,q_T,\eps,\tau/\omega)
\,.\end{align}
We can then apply the zero-bin subtraction to subtract overlap with the soft function, see \sec{zero_bin}, which for the exponential regulator is equivalent to dividing by the soft function in Fourier space~\cite{Luo:2019hmp}.
This in turn completes the manipulations that forced us the introduce the label $^{\rm naive}$ before their execution.
Next, we can apply the usual UV and IR counterterms to obtain the renormalized matching kernel as
\begin{equation} \label{eq:I_ren_qT}
 \tilde \cI_{ij}(x,b_T,\mu,\nu/\omega)
 = \sum_{j'} \Gamma_{jj'}(z, \eps) \otimes_z \tilde Z_B^i(\eps,\mu,\nu/\omega) \hat Z_{\as}(\mu,\eps)
   \frac{\tilde \cI_{ij'}^{\rm naive}(z,b_T,\eps,\tau/\omega)}{\tilde S(b_T,\eps,\tau)}
\,,\end{equation}
where following \refcite{Li:2016ctv} we identify the rapidity renormalization scale as $\nu = 1/\tau$.
The all-order structure of the beam function counter term $\tilde Z_B^i$ can be predicted from the beam function RGE, which we show in detail in \refcite{Ebert:2020yqt}.

We have implemented the described procedure at NLO through $\cO(\eps^4)$ and at NNLO through $\cO(\eps^2)$, as required for the calculation of the three-loop beam function.
We use the collinear limit of the cross sections for Higgs and Drell-Yan production to extract the gluon and quark beam functions, respectively. Since the bare soft function required in \eq{I_ren_qT} has not been published beyond NLO, we have similarly calculated it from the soft limit of the cross section.
Our bare results agree with those of \refscite{Luo:2019hmp,Luo:2019bmw}, and the renormalized beam functions also agree with \refcite{Luebbert:2016itl}.
We provide these beam functions, as well as the bare two-loop soft function, in ancillary files.

%% file: Chapters/Rapidity.tex
%%%%%%%%%%%%%%%%%%%%%%%%%%%%%%%%%%%%%%%%%%%%%%%%%%%%%%%%%%%%%%%%%%%%%%%%%%%%%%%%
\section{Collinear expansion of rapidity distributions}
\label{sec:rapidity}
%%%%%%%%%%%%%%%%%%%%%%%%%%%%%%%%%%%%%%%%%%%%%%%%%%%%%%%%%%%%%%%%%%%%%%%%%%%%%%%%
Computing analytic coefficient functions at high orders is a complicated task, and finding suitable approximations can be vital.
Here we demonstrate that our expansion techniques have the potential to approximate the rapidity spectrum of color neutral hard probes.
We perform a computation of the first two terms in the collinear expansion of the rapidity distribution of the Higgs boson produced via gluon fusion at NNLO.
This application also demonstrates that our technique allows one to relatively easily obtain predictions beyond leading power of the kinematic expansion.

The required partonic matrix elements were calculated exactly in \refcite{Gehrmann_De_Ridder_2012}, and the differential distribution was obtained for example in \refcite{Anastasiou2005}.
Currently, this observable is known at N$^3$LO computed via a threshold expansion~\cite{Dulat:2018bfe} and via an approximate differential computation~\cite{Cieri:2018oms}.
The exact computation of the partonic coefficient function is still elusive due to its extreme difficulty, and a collinear expansion of the same could provide a useful ingredient in future phenomenological studies.

We integrate out the degrees of freedom of the top quark and work in an effective theory that couples the Higgs boson directly to gluons~\cite{Inami1983,Shifman1978,Spiridonov:1988md,Wilczek1977,Chetyrkin:1997un,Schroder:2005hy,Chetyrkin:2005ia,Kramer:1996iq,Kniehl:2006bg}. 
We generate all required Feynman diagrams with QGRAF~\cite{Nogueira_1993} and perform their collinear expansion up to the second term as illustrated in \sec{collinear_expansion}.
To be precise, we only expand real emissions in the collinear limit, but keep all virtual loops exact. This differs from the calculation of beam functions presented in \sec{beamfunctions}, where a strict collinear limit was enforced by also expanding virtual loops in the collinear limit.
We then employ IBP identities~\cite{Chetyrkin:1981qh,Tkachov:1981wb} to reduce the expanded diagrams to master integrals, which we then compute using the framework of reverse unitarity~\cite{Anastasiou2003,Anastasiou:2002qz,Anastasiou:2003yy,Anastasiou2005,Anastasiou2004a} and the method of differential equations~\cite{Kotikov:1990kg,Kotikov:1991hm,Kotikov:1991pm,Henn:2013pwa,Gehrmann:1999as}.
With this we obtain the bare partonic coefficient function
\beq
\frac{\df\eta_{ij}}{\df Q^2 \df w_1 \df w_2 \df x}\Big|_{w_2 \sim \lambda^2}
\eeq
expanded up to the second term in $\omega_2$. 
Next, we perform a variable transformation from $(\omega_1,\omega_2)\rightarrow (z_1,z_2)$ via \eq{xidef} and \eqref{eq:partcoef_special}, and replace the variable $x$ by $\xi$ via
\beq
x=\frac{\xi (z_1+z_2){}^2}{\bigl[\xi z_1 (1-z_2) + z_2 (1 + z_1)\bigr] \bigl[\xi z_2 (1-z_1) + z_1(1+z_2)\bigr]}.
\eeq
The expansion in  $\omega_2$ is comparable to an expansion in $\bar z_2=1-z_2$, as can be seen by applying the rescaling transformation of \eq{modes}. We find
\beq
\bar z_1=1-z_1=\omega_1 +\mathcal{O}(\lambda^2),\hspace{1cm} \bar z_2=\omega_2 \frac{2-\omega_1(1+x)}{2(1-\omega_1)}+\mathcal{O}(\lambda^4).
\eeq
Introducing the variable $\xi$ has the advantage that its integration domain is independent from $z_1$ and $z_2$ and ranges from 0 to 1.
Next, we expand the partonic coefficient function after this change of variables up the second power in $\bar z_2$ and integrate over $\xi$.
The result is an approximation for the partonic coefficient function of \eq{partcoef_special} with the observable integrated out.
We perform UV renormalisation and combine our partonic matrix elements with collinear counter terms in order to obtain a finite partonic coefficient function through NNLO. 

Obtaining the equivalent expansion in $\bar z_1$ can easily be done by simply relabelling the variables, $\bar z_1\leftrightarrow \bar z_2$.
With this we obtain the following approximation for the full renormalized partonic coefficient function,
\begin{align} \label{eq:approxeta}
 \frac{\df\eta^{R,\,\text{approx.}}_{ij}(z_1,z_2) }{\df Q^2 \df Y} &
=\frac{\df\eta^{R}_{ij}}{\df Q^2 \df Y} \Big|_{\bar z_2 \sim \lambda^2}
+\frac{\df\eta^{R}_{ij}}{\df Q^2 \df Y} \Big|_{\bar z_1 \sim \lambda^2}
-\frac{\df\eta^{R}_{ij}}{\df Q^2 \df Y} \Big|_{\bar z_{1,2} \sim \lambda^2}
\quad
 + \cO(\lambda^2)
\,.\end{align}
The last term in the above equation removes the overlap in the two expansions. 
The hadronic cross section expanded to this order is then obtained by inserting \eq{approxeta} into \eq{sigma_hadr_finite},
\beq \label{eq:approx_sigma}
\frac{\df \sigma(x_1^B,x_2^B)}{ \df Q^2 \df Y} =
\tau \sigma_0 \sum_{i,j} f_i^R(x_1^B) \otimes_{x_1^B} \ \frac{\df\eta^{R,\,\text{approx.}}_{ij}(x_1^B,x_2^B) }{\df Q^2 \df Y}  \otimes_{x_2^B} f_j^R(x_2^B),
\eeq
Note that the leading-power limit of \eqs{approxeta}{approx_sigma} precisely correspond to the leading-power generalized threshold factorization theorem of \refcite{Lustermans:2019cau}, cf.~their eqs.~(17) and (18).

\begin{figure*}
\centering
\includegraphics[width=0.95\textwidth]{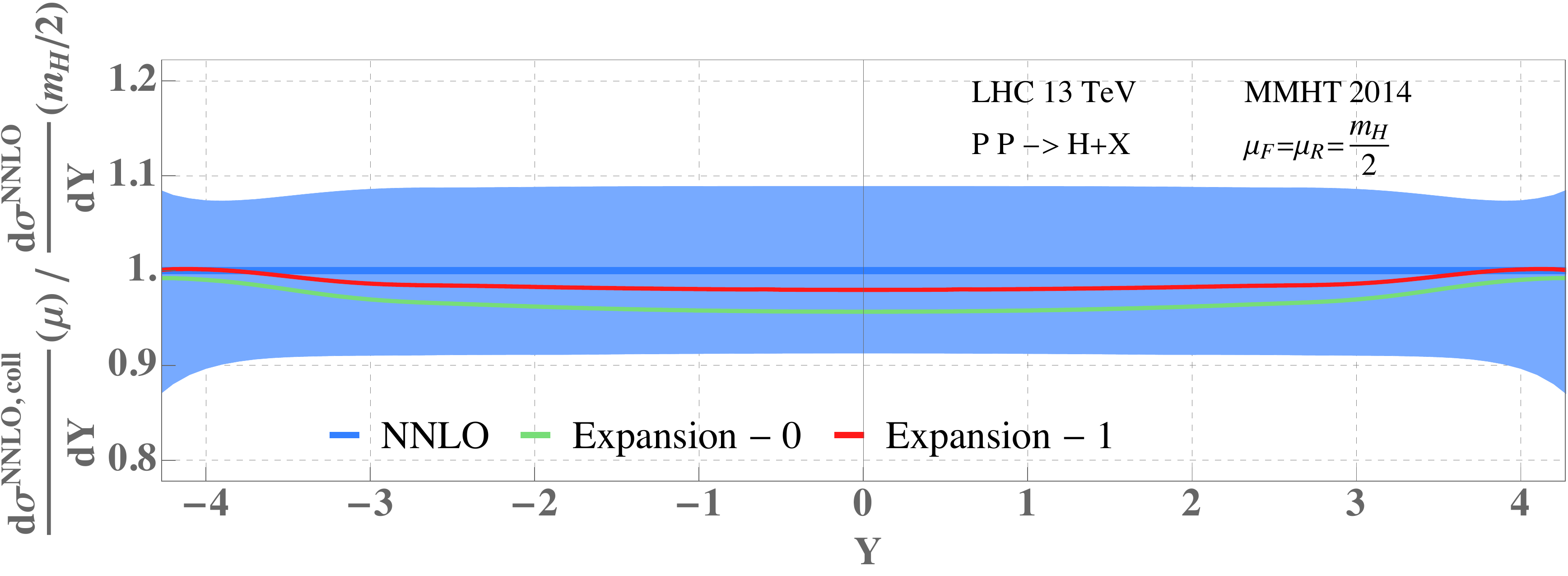}
\caption{Comparison of the Higgs boson rapidity distribution in gluon fusion obtained with a collinear expansion, normalized to the exact results of \refcite{Dulat:2017aa}.}
\label{fig:rap}
\end{figure*}

We have implemented the approximate partonic coefficient function in \eq{sigma_hadr_finite} in a private C++ code. Note that we only expanded the NNLO correction to the partonic coefficient, but keep the lower orders exact.
To illustrate our results numerically, we evaluate \eq{approx_sigma} for the LHC with a center-of-mass energy of $13$ TeV using the MMHT14 parton distribution functions~\cite{Harland-Lang:2014zoa}.
Figure~\ref{fig:rap} shows the rapidity distribution obtained with this collinear expansion normalized to the exact results obtained from \refcite{Dulat:2017aa}.
The green line shows our result using only the first term in the collinear expansion, while the red line shows the result including also the second term in the collinear expansion.
The blue band in the figure represents the variation of the cross section under a variation of the factorization and renormalisation scale by factor of two around their central values $\mu_F=\mu_R=m_H/2$.
We observe that the collinear expansion approximates the shape of the rapidity spectrum quite well, in particular towards large values of $|Y|$.
This is kinematically expected, as large rapidities enforce all final-state radiation to be collinear to the corresponding incoming parton, such that the collinear expansion is in fact the correct kinematic limit, see also~\refcite{Lustermans:2019cau}.
In addition, including the second-order term in the expansion clearly improves the results, illustrating that the collinear expansion indeed can be used to produce systematically improvable approximations of key collider physics observables.

The computation of the expanded partonic coefficient functions was greatly simplified compared to the computation of the exact result obtained e.g.\ in \refcite{Dulat:2017aa}.
Explicitly, the complexity of the analytic formulae is greatly reduced, and the function space required to express the coefficient function is much simpler.
We expect that a similarly drastic simplification will also occur when applying our method at N$^3$LO, which is a natural application of this research.

%% file: Chapters/Conclusion.tex
%%%%%%%%%%%%%%%%%%%%%%%%%%%%%%%%%%%%%%%%%%%%%%%%%%%%%%%%%%%%%%%%%%%%%%%%%%%%%%%%
\FloatBarrier
\section{Conclusions}
\label{sec:conclusions}
%%%%%%%%%%%%%%%%%%%%%%%%%%%%%%%%%%%%%%%%%%%%%%%%%%%%%%%%%%%%%%%%%%%%%%%%%%%%%%%%

We have developed a method to efficiently expand differential cross sections for the production of colorless final states in hadron collisions around the particular kinematic limit that all hadronic final-state radiation becomes collinear to one of the colliding hadrons. This yields a generalized power expansion in a power counting parameter $\lambda$ characterizing this limit.

A key feature of our method is that the expansion is systematically improvable, as it allows to compute to arbitrary order in the power counting parameter $\lambda$.
Furthermore, $\lambda$ is treated as a purely symbolic power counting parameter agnostic of the actual observable. This greatly simplifies the expansion, as it can be carried out at the integrand level, i.e. before any phase space or loop integrations are carried out.
Subsequently, carrying out phase space and loop integrals is greatly facilitated as integrands become simpler as a result of the expansion.
Moreover, the expanded integrands have again a diagrammatic nature very much like the original Feynman integrands they were derived from.
This observation makes it manifest that widely used and powerful loop integration techniques like IBP relations and the method of differential equations are applicable to the coefficients of the collinear expansion.
We also stress that the basic functions (the so-called master integrals) required in the computation of higher orders in the expansion are already obtained in the lowest few nontrivial orders of the expansion.

Our method also sheds light on the connection between the collinear limit of hadronic cross sections and factorization theorems derived in SCET. The latter include so-called beam functions, universal quantities defined as hadronic matrix elements of collinear fields in SCET, which can be related to standard light-cone PDFs through convolutions with perturbative matching kernels. We have shown that these kernels are precisely given by the first term in a strict collinear expansion of hadronic cross sections.
As a first application of this, we reproduced the matching kernels for the $N$-jettiness and $q_T$ beam functions at NNLO from a collinear expansion of the NNLO cross sections for the Drell-Yan process and for Higgs boson production in gluon fusion. The analytic results of this computation are provided as ancillary files together with the arXiv submission of this article.

As another application of the collinear expansion, we have demonstrated its usefulness to efficiently calculate approximate hadron collider cross sections.
By combining the collinear expansion with the limit where one partonic momentum fraction becomes equal to its Born value, $x_i\to x_i^B$, 
we obtained the first two terms in the collinear expansion of the rapidity distribution of a Higgs boson produced in gluon fusion through NNLO in QCD perturbation theory.
This example illustrates not only that key collider observables can be approximated with high accuracy using our technique, but also that results beyond the leading power can be easily obtained.

In summary, the method of collinear expansions is a great tool to study the infrared limit of QCD. At leading power in the collinear expansion, it provides access to the universal beam functions governing the collinear limit, which we employ to calculate the $\Tau_N$ and $q_T$ beam functions at N$^3$LO in two companion papers~\cite{Ebert:2020unb,Ebert:2020yqt}.
We also believe that the collinear expansions will similarly shed light on the universal structure of hadron collision processes beyond the leading power. Finally, it provides a powerful tool to achieve cutting-edge phenomenological predictions at very high orders in perturbation theory.

%%%%%%%%%%%%%%%%%%%%%%%%%%%%%%%%%%%%%%%%%%%%%%%%%%%%%%%%%%%%%%%%%%%%%%%%%%%%%%%%
\acknowledgments
We thank Martin Beneke, Johannes Michel, Ian Moult, Iain Stewart, Frank Tackmann and Hua Xing Zhu for useful discussions.
This work was supported by the Office of Nuclear Physics of the U.S.\ Department of Energy under Contract No.\ DE-SC0011090 and DE-AC02-76SF00515.
M.E.\ is also supported by the Alexander von Humboldt Foundation through a Feodor Lynen Research Fellowship,
and B.M.\ was also supported by a Pappalardo fellowship.
%%%%%%%%%%%%%%%%%%%%%%%%%%%%%%%%%%%%%%%%%%%%%%%%%%%%%%%%%%%%%%%%%%%%%%%%%%%%%%%%